\documentclass[journal,draftcls,onecolumn,12pt,twoside]{IEEEtranTCOM}
\normalsize

\usepackage{balance}
\usepackage{arydshln}
\usepackage{multirow}

\usepackage{graphicx}
\usepackage{rotating}
\usepackage{tikz}
\usepackage{amssymb}
\usepackage{epsfig}
\usepackage{subfigure}
\usepackage{epstopdf}
\usepackage{algorithm}
\usepackage{algorithmic}
\usepackage{amsmath}
\usepackage{amsfonts}
\usepackage{dsfont}
\usepackage{url}
\usepackage{chemarr}
\usepackage{chemarrow}
\usepackage{bbold}
\usepackage[version=3]{mhchem}
\usepackage{mathabx}

\newif\ifusetables
\usetablesfalse 

\ifCLASSINFOpdf
\else
\fi

\begin{document}
%
\title{\textcolor{black}{Generalized} Solution for the Demodulation of Reaction Shift Keying Signals in Molecular Communication Networks}
%
%
%
\author{Hamdan Awan,~\IEEEmembership{Student Member,~IEEE,}
        and Chun Tung Chou~\IEEEmembership{Member,~IEEE,}
\thanks{H. Awan and C. T. Chou are with the School
of Computer Science and Engineering, The University of New South Wales, Sydney,
NSW 2052, Australia. e-mail: (hawan,ctchou@cse.unsw.edu.au)}}

%
%

\markboth{IEEE Transactions on Communications}%
{Submitted paper}
%



\maketitle
\begin{abstract}
This paper considers a diffusion-based molecular communication system where the transmitter uses Reaction Shift Keying (RSK) as the modulation scheme. We focus on the demodulation of RSK signal at the receiver. The receiver consists of a front-end molecular circuit and a back-end demodulator. The front-end molecular circuit is a set of chemical reactions consisting of multiple chemical species. The optimal demodulator computes the posteriori probability of the transmitted symbols given the history of the observation. The derivation of the optimal demodulator requires the solution to a specific Bayesian filtering problem. The solution to this Bayesian filtering problem had been derived for a few specific molecular circuits and specific choice(s) of observed chemical species. The derivation of such solution is also lengthy. The key contribution of this paper is to present a general solution to this Bayesian filtering problem which can be
applied to any molecular circuit and any choice of observed species. 
\end{abstract}


%
\IEEEpeerreviewmaketitle

\section{Introduction}
 \label{sec:intro}
\IEEEPARstart{M}{olecular} communication \cite{Akyildiz:2008vt,Hiyama:2010jf,Nakano:2014fq} is a communication paradigm inspired by the communication between living cells and organisms. Molecular communication has many potential applications, e.g. health monitoring, therapeutics etc \cite{Atakan:2012ej,Nakano:2014fq}. In this paper, we deal with diffusion-based molecular communication \cite{Pierobon:2014iu} where the signalling molecules diffuse freely in the transmission medium between the transmitter and the receiver. 

Modulation and demodulation are important building blocks of any communication system. For diffusion-based molecular communication, many different modulation schemes have been proposed: Molecule Shift Keying (MSK) \cite{ShahMohammadian:2012iu,Kuran:2011tg}, Frequency Shift Keying (FSK) \cite{Chou:2012ug}, Pulse Position Modulation (PPM) \cite{Kadloor:ua}, Concentration Shift Keying (CSK) \cite{Atakan:2010bj,Mahfuz:2011te} and token communication \cite{Rose:hn}. Recently we proposed a new modulation scheme called Reaction Shift Keying (RSK) \cite{7208820,Awan:2015:IRM:2800795.2800798}. RSK is inspired by studies in intra-cellular and inter-cellular communication in biology which show that living cells use temporal code or signalling dynamics for communication \cite{Kubota:2012fe,purvis2013encoding}. A temporal code is a signal consisting of varying number (or concentration) of signalling molecules over time. It is known from these studies in biology that the time-varying signals can be produced by networks of chemical reactions, which are also known as molecular circuits \cite{Kubota:2012fe,purvis2013encoding}. With this background in mind, RSK uses different chemical reactions to generate different time-varying concentration of signalling molecules to represent different transmission symbols.

The receiver uses a molecular circuit to process the incoming signal. When the signalling molecules arrive at the receiver, they react with the chemicals in the receiver molecular circuit to produce one or more types of output molecules. The counts of these output molecules over time will be the output signals which contain information on the symbol sent by the transmitter. In our earlier work, we have studied two specific choices of molecular circuits at the receiver: ligand-receptor binding in \cite{7208820} and a receptor with two binding sites in \cite{Awan:2015:IRM:2800795.2800798} and a protein kinase circuit \cite{Awan2016}. For both pieces of work, we calculate the posteriori probability that a particular symbol has been transmitted. We find that the logarithm of the posteriori probability (up to a constant) can be obtained from the output of an analog filter and the key contribution of the earlier work is to derive these filters. A key step in deriving these filters is to solve a Bayesian filtering problem. This Bayesian filtering problem has to be solved for each type of molecular circuit and each choice of output molecules. The mathematical derivation of the solution to the Bayesian filtering problem is also very lengthy. 

\textcolor{black} {
The key contribution of this paper is to present a general solution to the Bayesian filtering problem. An advantage of the proposed method is that it can be used with any receiver molecular circuit and any choice of output molecules. Another advantage is that it by-passes the need to go through long mathematical derivation. The solution provides a method to write dow the solution to the filtering problem directly. To the best of our knowledge, this generalized solution does not appear to have been studied before. 
}

The rest of the paper is organised as follows. Section \ref{sec:related} discusses the related work. Section \ref{sec:model} presents the end-to-end model and solution to the Bayesian filtering problem using the earlier approach, while Section \ref{133} presents the general solution. Section \ref{sec:eval} present numerical examples and Section \ref{22} concludes the paper.

\section{Related work} 
\label{sec:related}
The interest of research community in molecular communication is on the rise as shown by recent surveys \cite{Akyildiz:2008vt,Hiyama:2010jf,Nakano:2012dv,Nakano:2014fq,Farsad:2014vl}. 

On the transmitter side different modulation schemes have been proposed in literature as mentioned in Section \ref{sec:intro}. These schemes also use different signalling molecules emission pattern at the transmitter, e.g. impulse \cite{Mahfuz:2011kg}, Poisson process \cite{mosayebi2014receivers}. However, this paper focuses on RSK where the transmitter uses different chemical reactions to generate different emission patterns to represent different symbols \cite{7208820,Awan:2015:IRM:2800795.2800798}. 

For the receiver side, different receiver designs have been proposed in literature for molecular communication systems, e.g. \cite{Noel:2014fv,Noel:2014hu,Kilinc:2013by,Mahfuz:2014vs,Meng:2014hh}. Similarly different demodulation schemes for molecular communication systems are presented in \cite{Noel:2014hu,Mahfuz:2014vs}. A common idea in these papers is that discrete-time samples of the number of output molecules are used to compute the likelihood of the transmitted symbol. However, the demodulation of RSK uses continuous-time signals \cite{7208820,Awan:2015:IRM:2800795.2800798} where the processing of such continuous-time signals requires an analog filter. We further shown in our earlier work \cite{7208820,7206607} that information processing using the uniformly sampled version of the signals generally will result in information loss. Similar conclusion was also arrived in \cite{Mahdavifar15} from an information theoretic analysis on capacity. 

\textcolor{black}{
The demodulation of RSK signals have previously been considered in \cite{7208820,Awan:2015:IRM:2800795.2800798}. However, each of these works considers only a specific choice of molecular circuit. Instead the results of this paper are general. The general algorithm can be applied to {\bf any} receiver molecular circuit and for {\bf any} choice of measurements.
}

An alternative way of designing receivers for molecular communication is by using molecular circuits, see \cite{Chou:2014jca,Chou:hf} for example. Various aspects of receiver molecular circuits have been studied in the literature. We will discuss two aspects here: capacity and noise properties. The information transmission capacity of a number of types of linear receiver molecular circuits is compared in \cite{Chou:2014jca}. The capacity analysis for molecular communication based on ligand receptor binding has been presented in \cite{Einolghozati:2011ge,Einolghozati:2011cj}. The capacity of these systems in the continuous-time is presented in \cite{Thomas:2014tf}. The noise properties of ligand-receptor binding type of receivers is studied in \cite{Pierobon:2014iu,Pierobon:2011ve}. All the above papers assume that the receiver is a ligand-receptor binding process with only two reactions: binding and unbinding. However, in this paper, we propose a methodology that can be used for any molecular circuit. 

Different models have been used in molecular communication literature to model the transmission medium. The papers \cite{Mahfuz:2011kg,mosayebi2014receivers} assume that medium is continuous while in this paper, as well as in our previous work \cite{7208820,Awan:2015:IRM:2800795.2800798}, we assume that the medium is divided into cubic voxels. The use of voxels allows us to model the end-to-end communication system using reaction diffusion master equation (RDME) \cite{Chou:rdmex_tnb,Chou:rdmex_nc,7208820}, which is a continuous-time Markov Process (CTMP). An alternative end-to-end model appears in \cite{Pierobon:2011ve,Pierobon:2011vr} which is based on particle tracking. An advantage of the RDME approach is that we can use the Markovian properties to analyse molecular communication \cite{7206607,7208820}.

\begin{figure}
\begin{center}
\includegraphics[page=1,trim=0cm 0cm 0cm 0cm ,clip=true, width=0.45\columnwidth]{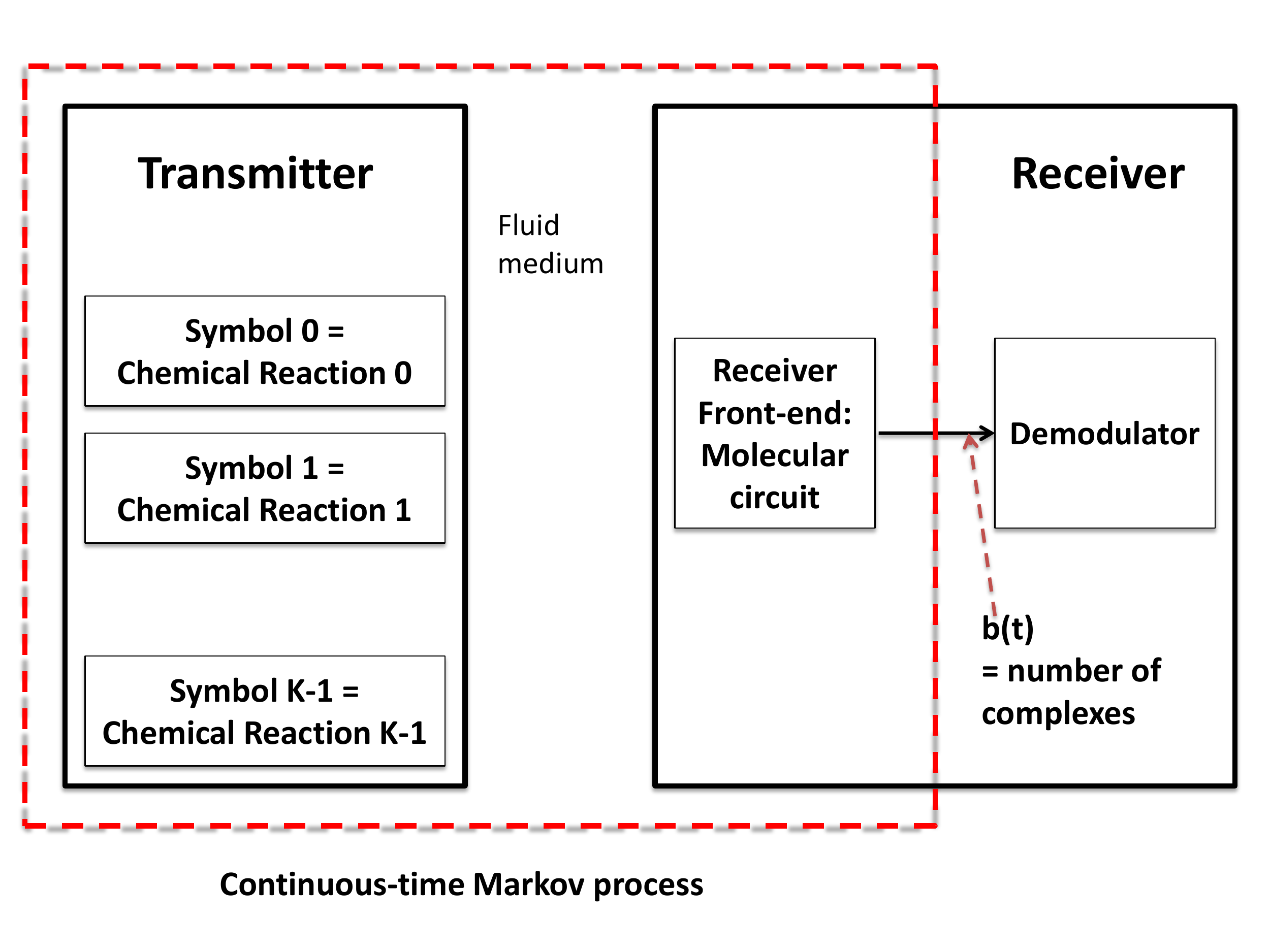}
\caption{System Overview }
\label{fig:overall}
\end{center}
\end{figure}

\section{Summary of the Current Approach} 
\label{sec:model} 
This section presents the current approach of deriving the optimal demodulation filter for RSK modulation, which appeared in \cite{7208820,Awan:2015:IRM:2800795.2800798}. In particular, this section summarises the method used in our previous conference publication \cite{Awan:2015:IRM:2800795.2800798} when it is applied to a ligand-receptor binding process with two binding sites. We will show in Section \ref{133} how the same result can be derived using a general algorithm. \textcolor{black}{We want to stress that the aim of this paper is to provide a general algorithm to derive the solution to an optimal Bayesian filtering problem which is an intermediate step needed to derive the demodulation filters for RSK modulation. In particular, the general algorithm will be applicable to any receiver molecular circuit and for any choice of measurements. The receiver circuit of ligand-receptor with two binding sites is used to illustrate the applicability of the general algorithm. We choose this particular receiver circuit because it is a small circuit that can be used to illustrate all the important features of the general algorithm.}

This section is organised as follows. Section \ref{sec:sys:model} presents the system assumptions and the resulting end-to-end model is presented in 
Section \ref{123}. An overview of the maximum a posteriori (MAP) demodulation framework is discussed in Section \ref{sec:MAP}. The derivation of the MAP demodulator requires the solution to an optimal Bayesian filtering problem and this is discussed in Section \ref{24}. Finally, the demodulation filters are derived in Section \ref{2222}. 
\subsection{System Model and Assumptions}
\label{sec:sys:model}
In this section we present the modelling assumptions of a diffusion based molecular communication system that uses the RSK modulation scheme. \figurename~\ref{fig:overall} depicts an overview of the system, which consists of a transmitter, a fluidic transmission medium and a receiver. We assume that the transmitter and receiver are synchronized. The transmitter generates signalling molecules which are diffuse freely in the transmission medium. The receiver consists of two parts: the front-end is a molecular circuit and the back-end is the demodulator. When the signalling molecules reach the receiver molecular circuit, they react with the chemical species in the circuit to generate one or more types of output molecules. The history of the number of these output molecules over time is the input to the demodulation block. 

A key modelling assumption that we make is that the medium is discretized in voxels while time is continuous. This allows us to model the end-to-end system using RDME, which is a type of CTMP, which describes the time evolution of the number of signalling molecules in the transmitter, medium and receiver. We assume the system uses one type of ligands or signalling molecules $S$ for transmitting the information. Next we present the models for propagation medium, transmitter and receiver.

\begin{figure}
\begin{center}
\includegraphics[page=8,trim=0cm 0cm 0cm 0cm ,clip=true, width=0.45\columnwidth]{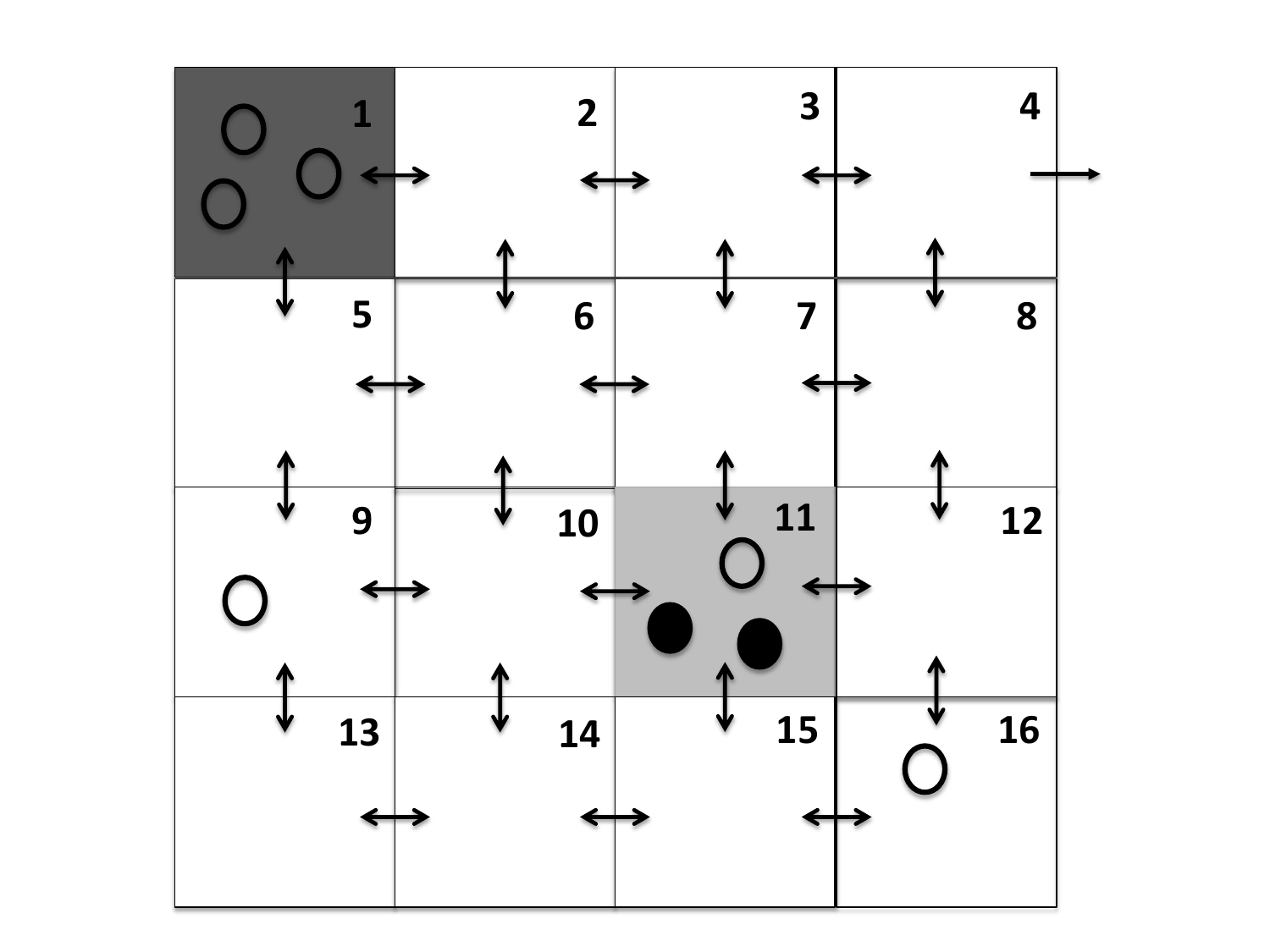}
\caption{A Model of Molecular Communication Network.}
\label{fig:model}
\end{center}
\end{figure}

\subsubsection{Medium of Propagation} 
\label{sec:model:prop} 
Signalling molecules diffuse from transmitter to receiver through a propagation medium. We assume the medium of propagation is a three dimensional space of dimension $X \times Y \times Z$ where $X$, $Y$ and $Z$ are an integral multiple of length $W$, i.e. there exist integers $N_x$, $N_y$ and $N_z$ such that $X = N_x W$, $Y = N_y W$ and $Z = N_z W$. The entire space is therefore divided into $N_x \times N_y \times N_z$ cubic voxels of volume $W^3$. Fig.~\ref{fig:model} shows an example with $N_x$ $=$ $N_y$ $=$ $4$ and $N_z$  $=$ $1$; note that the figure should be viewed as a projection onto the $x$-$y$ plane. We will index the voxels by using a single index  $\xi \in [1,N_x N_y N_z]$, e.g. in Fig.~\ref{fig:model}, the integer in the top-right corner of the voxel indicates the index of that voxel.

Diffusion can be modelled by the movement of molecules between neighbouring voxels. For example in \figurename~\ref{fig:model}, signalling molecules can freely diffuse from Voxel $1$ to its two neighbouring voxels, which are Voxels 2 and 5. Similarly signalling molecules in Voxel $2$ can move to either Voxel $1$, $3$ or $6$. The arrows in \figurename~\ref{fig:model} show the direction of diffusion. We assume the diffusion coefficient $D$ of signalling molecules is constant in the medium. By discretising the diffusion equation, it can be shown that the diffusion of molecules from one voxel to another occurs at a mean rate $d = \frac{D}{W^2}$ \cite{Gardiner}. Furthermore, the probability that a signalling molecule moves from a voxel to a neighbouring voxel  within an infinitesimal time interval $\Delta t$ is $d \Delta t$. Lastly, our model can be used to model two types of boundary conditions: reflecting boundary condition where signalling molecules are not allowed to leave the medium; or, absorbing boundary condition where signalling molecules may leave the medium forever, e.g. the single headed arrow in Voxel 4 in Fig.~\ref{fig:model} shows that signalling molecules may leave the medium. 

\subsubsection{Transmitter Model} 
\label{sec:model:transmitter}
We assume in this paper that the transmitter occupies one voxel but generalisation to a multi-voxel transmitter is straightforward. In \figurename{}~\ref{fig:model}, we assume that the transmitter $T$ occupies Voxel $1$. The transmitter uses RSK modulation scheme with $K$ transmission symbols indexed by $s = 0,1,..,K-1$ as shown in \figurename{} \ref{fig:overall}. Each symbol is associated with a specific emission pattern $u_s(t)$ which is produced by a set of chemical reactions. These reactions take place in the transmitter voxel. The role of this emission pattern is similar to a transmitted signal in communication. To explain the meaning of emission pattern we consider following example. Suppose the emission pattern $u_s(t)$ represents symbol $s$ such that $u_s(3) = u_s(6) = 1$ and  $u_s(t) = 0$  for all other values of $t$. This means that for symbol $s$, a molecule is emitted at each of the times $t = 3$ and $t=6$ while no emissions take place at other times. In RSK, the emission patterns for the $K$ different symbols are produced by $K$ different sets of chemical reactions. An example chemical reaction occurring inside living cells is \cite{Tkacik:2011jr}:
\begin{align}
\cee{
RNA &  ->[\kappa] RNA +  S \label{eq:t1:1}
} 
\end{align}
where ribonucleic acid (RNA) produces the molecule $S$. This chemical reaction can be modelled by a Poisson process where molecules $S$ are produced at a mean rate equals to $\kappa$ times the number of RNA molecules, denoted by $n_{\rm RNA}$ \cite{Shahrezaei:2008bp}. The mean emission pattern produced by Reaction \eqref{eq:t1:1} is ${\bf E}[u(t)] = \kappa n_{\rm RNA}$. In general, it was shown in \cite{Gillespie:1992tq} that chemical reactions can be modelled by a CTMP, therefore in this paper we use CTMP to model the reactions in the transmitter. Note that chemical reactions are stochastic, so a set of chemical reactions can generate an infinite number of emission patterns but with different probabilities. In this paper, we will not specify the chemical reactions in the transmitter because the optimal demodulation filters does not explicitly depend on these chemical reactions.

\subsubsection{Receiver}
\label{sec:model:rec} 
We assume in this paper that the receiver occupies a single voxel but generalisation to multi-voxel receiver is straightforward. For example, in Fig.~\ref{fig:model}, the receiver $R$ occupies Voxel $11$. The receiver is further divided into two blocks: a molecular circuit at the front-end and the demodulator at the back-end as shown in \figurename~\ref{fig:overall}. In this section, we assume the front-end molecular circuit consists of receptors with two binding sites; we will consider general molecular circuits in Section \ref{133}. 

We use $E$ to denote an unbound receptor and we assume that there are $M$ receptors. These receptors are assumed to be fixed and homogeneously distributed in the receiver voxel. The receptor $E$ has two binding sites to which the signalling molecules $S$ can bind. The reactions at the receptors are:  
\begin{align}
\cee{
S + E &<=>C[\tilde{\lambda}_1][\mu_1] C_{[1]} \label{eq:mc1:r1}  \\
S + C_{[1]}  &<=>C[\tilde{\lambda}_2][\mu_2] C_{[2]} \label{eq:mc1:r2}  }
\end{align}
where $\tilde{\lambda}_1$, $\mu_1$, $\tilde{\lambda}_2$ and $\mu_2$ are reaction rate constants. In Reaction \eqref{eq:mc1:r1}, $E$ binds with a $S$ molecule to form the complex $C_{[1]}$ in the forward reaction. Similarly in the forward reaction in \eqref{eq:mc1:r2}, $C_{[1]}$ binds with another $S$ molecule to form the complex $C_{[2]}$. The number of complexes $ C_{[1]}$ and $ C_{[2]}$ at time $t$ is represented by $b_1(t)$ and $b_2(t)$ respectively. Since the total number of receptors is $M$, the number of unbound receptors is $M - b_1(t) - b_2(t)$. \textcolor{black}{Molecules with multiple binding sites have been studied in biology before. For example,  \cite{Libby:2007fw} studies the role of a molecule with multiple binding sites in the estimation of sugar concentration by bacteria. Our model in \eqref{eq:mc1:r1} and \eqref{eq:mc1:r2} are similar to the one in \cite{Libby:2007fw}. There is also a rich literature in biology on cooperativity, which studies the behaviour of molecules with multiple binding sites, see \cite{Keener} and the references therein.}

\textcolor{black}{
The reaction rate constants in \eqref{eq:mc1:r1} and \eqref{eq:mc1:r2} are based on concentration of reactants. Since the CTMP is based on molecular counts, we need to scale $\tilde{\lambda}_1$ and $\tilde{\lambda}_2$ by the volume of the voxels $W^3$. Specifically, we define $\lambda_1 = \frac{\tilde{\lambda}_1}{W^3}$ and $\lambda_2 = \frac{\tilde{\lambda}_2}{W^3}$ which will be used in the CTMP. For further explanation of this scaling, see \cite{Erban:2009us,7208820}. In the next section we present the end-to-end model for the complete system.
}

\subsection{End-to-End Model}
\label{123}
\label{sec:e2e}
In order to derive the optimal demodulator, we need an end-to-end model of system which includes the transmitter, receiver and medium of propagation. Since both diffusion and reaction can be modelled by CTMP \cite{Erban:2007we}, we choose to use CTMP (or RDME) as the end-to-end model. Note that we need one CTMP per transmission symbol (which corresponds to a set of chemical reactions.) The received signal may be affected by the noise in the system due to the reactions at transmitter and receiver, or due to diffusion in the medium. The advantage of using CTMP is that it includes all these noise sources in the end-to-end model.

Recall that the transmitter uses $K$ different sets of chemical reactions, we require one CTMP for each set of reactions. Given all these $K$ CTMP's have the same form, we will present a generic description that applies to all the symbols. The state of the CTMP for the end-to-end system consists of the counts of all chemical species in each voxel of the system. Mathematically, we write the state of the system as $(N(t),b_1(t),b_2(t))$ where $b_1(t)$ and $b_2(t)$ are the number of $C_{[1]}$ and $C_{[2]}$ molecules, as defined earlier. The vector $N(t)$ consists of the counts of all other chemical species (i.e. all chemical species with the exception of the receptors in the receiver voxel) in each voxel of the system. For example, $N(t)$ contains the counts of the chemical species in the transmitter voxels as well as the counts of the signalling molecules in all voxels. We assume that the $R$-th element of $N(t)$ is the count of the number of signalling molecules in the receiver voxel at time $t$, which is denoted as $n_R(t)$.

 A CTMP is specified by defining its state transition probabilities. State transition in this CTMP can be caused by reactions in the transmitter, diffusion of signalling molecules and reactions in the receiver. We will first consider the state transitions due to the reactions \eqref{eq:mc1:r1} and \eqref{eq:mc1:r2} in the molecular circuit in the receiver. 

If the forward reaction in \eqref{eq:mc1:r1} takes place between $[t,t+\Delta t)$, then a signalling molecule in the receiver voxel is consumed and the number of complex $C_{[1]}$ is increased by one. Also, this reaction occurs at a rate of $\lambda_1 n_R(t) (M - b_1(t) - b_2(t))$ by applying the law of mass action to \eqref{eq:mc1:r1}. In terms of state transition of the CTMP, we write:
\begin{align}
& {\bf P}[N(t+\Delta t) = N(t)-\mathbb{1}_R, b_1(t+\Delta t) = b_1(t) + 1, b_2(t+\Delta t) = b_2(t)| N(t), b_1(t),b_2(t)] \nonumber \\&  = \lambda_1 \; n_R(t)    \; (M-b_1(t)-b_2(t)) \; \Delta t
\label{eqn:tp:r1}
\end{align}

where $\mathbb{1}_R$ is a unit vector with a 1 in the $R$-th position and zeros at other positions. The condition $N(t+\Delta t) = N(t)-\mathbb{1}_R$ says that the number of signalling molecules in the receiver voxel has been decreased by one. The indicator function $\delta(.)$ is one when all the equalities within the function are true. Similarly, for the reverse reaction in \eqref{eq:mc1:r1}, we have: 
\begin{align}
& {\bf P}[N(t+\Delta t) = N(t)+\mathbb{1}_R, b_1(t+\Delta t) = b_1(t) - 1 ,  b_2(t+\Delta t)  =b_2(t)| N(t), b_1(t),b_2(t)] \nonumber \\&
 = \mu_1 \; b_1(t) \; \Delta t
\label{eqn:tp:r2} 
\end{align}

In the same way, the state transitions due to the reactions in \eqref{eq:mc1:r2} are: 
\begin{align}
& {\bf P}[N(t+\Delta t) = N(t)-\mathbb{1}_R, b_1(t+\Delta t) = b_1(t) - 1, b_2(t+\Delta t) = b_2(t) + 1| N(t), b_1(t),b_2(t)]  \nonumber \\&  = \lambda_2 b_1(t) \; n_R(t) \; \Delta t
\label{eqn:tp:r3}
\end{align}
\begin{align}
& {\bf P}[N(t+\Delta t) = N(t)+\mathbb{1}_R, b_1(t+\Delta t) = b_1(t) + 1 ,  b_2(t+\Delta t) = b_2(t) - 1| N(t), b_1(t),b_2(t)] \nonumber \\&
= \mu_2 \; b_2(t) \; \Delta t
\label{eqn:tp:r4} 
\end{align}

We now specify the transition probabilities when $b_1(t)$ and $b_2(t)$ remain unchanged. These transitions are caused by either a reaction in the transmitter or diffusion of signalling molecules between neighbouring voxels. Let $\eta_i, \eta_j$ be two valid $N(t)$ vectors. For $\eta_i \neq \eta_j$, we write 
\begin{align}
 & {\bf P}[N(t+\Delta t) = \eta_i, b_1(t+\Delta t) = b_1(t),  b_2(t+\Delta t) = b_2(t) |  N(t) = \eta_j, b_1(t), b_2(t)]  = d_{ij} \; \Delta t 
\label{eqn:tp:eta}
\end{align}
where $d_{ij}$ is the transition rate when $N(t)$ changes from $\eta_j$ to $\eta_i$. Since this transition is due to either a reaction in the transmitter or diffusion, $d_{ij}$ is independent of the number of the two complexes. Depending on the type of transition, the value of $d_{ij}$ can depend on the reaction constants in the transmitter, diffusion rate and some states of $\eta_j$. 
 
The main advantage of using Equation \eqref{eqn:tp:eta} is that it allows us a cleaner abstraction to solve the Bayesian filtering problem when deriving the MAP demodulator. We also remark that we will not specify the exact expression of $d_{ij}$ because $d_{ij}$'s do not appear explicitly in the demodulator. 

Equations \eqref{eqn:tp:r1} to \eqref{eqn:tp:eta} specify all the possible state transitions. The probability of no state transition is therefore:
\begin{align}
 & {\bf P}[N(t+\Delta t) = N(t), b_1(t+\Delta t) = b_1(t),  b_2(t+\Delta t) = b_2(t) | N(t) = \eta_j, b_1(t), b_2(t)]  \nonumber \\ 
& =  1 - [\sum_{i \neq j} d_{ij}  - \lambda_1 \; n_R(t) \; (M-b_1(t)-b_2(t)) -   
 \mu_1 b_1(t) - \lambda_2 b_1(t) n_R(t) - \mu_2 b_2(t) ]\; \Delta t  
\label{eqn:tp:no}
\end{align}

\subsection{The MAP framework} 
\label{sec:MAP} 
This section presents an overview of the MAP framework for the decoding of RSK signals. We assume that there is no inter-symbol interference (ISI) and we focus on the demodulation of one transmission symbol. There is no loss in generality because the same demodulator will be used whether ISI is present or not. The reader can refer to our earlier work \cite{7208820} on how ISI can be handled using decision feedback for ligand-receptor circuit (with one binding site) when the number of receptors is large. Note that it is still an open research problem on how to handle ISI for general molecular circuits. The difficulty comes from nonlinearity in chemical reactions. 

In the MAP framework, the key idea is to compute the posteriori probability of the transmitter having sent symbol $s$ given the observations. In our set up in Fig.~\ref{fig:overall}, the observations available to the demodulator are the counts of certain chemical species in the molecular circuits. We will consider three different choices of observations: (a) Only the number of complexes $ C_{[1]}$ over time, i.e $b_1(t)$, is available to the demodulator; (b) Number of complexes $ C_{[2]}$, i.e $b_2(t)$, only; (c) Number of both complexes, which is denoted by the vector signal $b_{A}(t) = (b_1(t),b_2(t))$ where the subscript $A$ is short for ``All". In general, we will denote the signal available to the demodulator as $b_{m}(t)$ where $m = 1, 2$ or $A$. 

The demodulator will make use of the continuous-time history of the counts of complex(es) as the input(s). We use ${\cal B}_m(t) = \{ b_m(\tau); 0 \leq \tau \leq t \}$ where $m=1,2,A$ to denote the continuous-time history of the input signal(s) up till time $t$. The aim of the MAP framework is to compute the probability ${\mathbf P}[s | {\cal B}_m(t)]$, which is the probability that the transmitter has sent symbol $s$ given the history ${\cal B}_m(t)$ up till time $t$. If the demodulator makes the decision at time $t$, then the estimated symbol $\hat{s}$ is: 
\begin{align}
\hat{s} = {\arg\max}_{s = 0, ..., K-1}  {\mathbf P}[s | {\cal B}_m(t)]\label{as}
\end{align}

We will now explain how the posteriori probability can be computed. Let us define $L_s(t) = \log( {\mathbf P}[s | {\cal B}_m(t)] )$ because it is easier to work with the logarithm of the probability. By using Bayes' rule, we have: 
\begin{align}
L_s(t+\Delta t) = & L_s(t) + \log (\mathbf{P}[b_m(t+\Delta t) | s, {\cal B}_m(t)])   - \log (\mathbf{P}[b_m(t+\Delta t) | {\cal B}_m(t)])
\label{eqn:logpp_prelim}
\end{align}
The last term in \eqref{eqn:logpp_prelim} does not depend on $s$, so it does not have to be calculated. The term $\mathbf{P}[b_m(t+\Delta t) | s, {\cal B}_m(t)]$ aims to predict the probability distribution of $b_m(t+\Delta t)$ from the history of observations is a Bayesian filtering problem. In Section \ref{24}, we will present the solution to the Bayesian filtering problem. Once the solution has been obtained, it can be substituted into \eqref{eqn:logpp_prelim} to obtain the demodulation filter, which will be presented in Section \ref{2222}.

\subsection{Solution of Bayesian Filtering Problem}
\label{24}
This section presents the expressions for $\mathbf{P}[b_m(t+\Delta t) | s, {\cal B}_m(t)]$ for three choices of inputs ($m = 1, 2, A$) to the demodulator. These expressions can be obtained by solving a Bayesian filtering problem where the system model is given by the CTMP in Section \ref{123}. Standard methods can be applied to solve the filtering problem but the derivation is lengthy, see \cite{7208820} for the case of ligand-receptor binding. In this paper, we will not present the derivation but we will simply state the results. We present the results for $m = 1$ and $m = 2$ first, and then followed by that for $m = A$.

\subsubsection {Solution for measuring $C_{[1]}$ only, i.e. $m = 1$}
\label{sec:bf:1or2} 
For $m = 1$, the probability $\mathbf{P}[b_m(t+\Delta t) | s, {\cal B}_m(t)]$ takes the following form: 
\begin{align}
& \mathbf{P}[b_1(t+\Delta t) | s, {\cal B}_1(t)]    
	=  \delta(b_1(t+\Delta t) = b_1(t) - 1) (Q_{1,1}+Q_{2,1}) +   \delta(b_1(t+\Delta t) = b_1(t) + 1)  \nonumber  \\ 
		    &  (Q_{3,1}+Q_{4,1}) +  \delta(b_1(t+\Delta t) = b_1(t)) Q_{5,1}   
	   \label{eqn:predictb}
\end{align}
where 
 \begin{align}
 &  Q_{1,1}=  \lambda_2  b_1(t) E[n_R(t) | s, {\cal B}_1(t)]  \label{eqn:q11} \Delta t\\
   &  Q_{2,1} =   \mu_1 b_1(t) \Delta t , Q_{3,1} = \lambda_1   E[(M- b_1(t)-b_2(t))n_R(t) | s, {\cal B}_1(t))] \Delta t   \nonumber\\
   & Q_{4,1} = \mu_2 E[b_2(t) | s, {\cal B}_1(t)] \Delta t,  Q_{5,1} =  1 -  (Q_{1,1} + Q_{2,1} + Q_{3,1} + Q_{4,1}) 
 	   \label{eqn:predictbx}
 \end{align}
  
Note that the probability $\mathbf{P}[b_1(t+\Delta t) | s, {\cal B}_1(t)]$ depends on whether $b_1(t+\Delta t)$ is one more, one less or equal to $b_1(t)$. The term $Q_{i,m}$ has two subscripts $i$ and $m$. The subscript $i$ varies from 1 to 4 for the four reactions causing a change in the number of $C_{[1]}$ moelcules whereas $i=5$ represents no change. The second subscript $m$ is used to indicate the choice of measurement.

\subsubsection {Solution for measuring $C_{[2]}$ only, i.e. $m = 2$}
\label{sec:bf:2} 
For $m = 2$, the probability $\mathbf{P}[b_m(t+\Delta t) | s, {\cal B}_m(t)]$ takes the following form:
\begin{align}
 & \mathbf{P}[b_2(t+\Delta t) | s, {\cal B}_2(t)]  
 	=  \delta(b_2(t+\Delta t) = b_2(t) + 1) Q_{1,2} +  \delta (b_2(t+\Delta t)= b_2(t) - 1) Q_{2,2}  + \nonumber  \\  
 	    & \delta(b_2(t+\Delta t) = b_2(t)) Q_{3,2} 
	   \label{eqn:predictbz}
\end{align}  
where
 \begin{align}
  & Q_{1,2} = \lambda_2   E[b_1(t)n_R(t) | s, {\cal B}_2(t)] \Delta t  ,  Q_{2,2} =   \mu_2 b_2(t) \Delta t ,  Q_{3,2} =  1 -  (Q_{2,1} +  Q_{2,2}) 
 	   \label{eqn:predictbx1}
 \end{align}

Note that the value of the probability $\mathbf{P}[b_2(t+\Delta t) | s, {\cal B}_2(t)]$ depends on whether $b_2(t+\Delta t)$ is one more, one less or equal to $b_2(t)$.  
 
\subsubsection {Solution for measuring both $C_{[1]}$ and $C_{[2]}$, i.e. $m = A$}
\label{sec:A}
For the case when the counts of both complexes are used as the inputs to the demodulator, we are interested to find the probability 
$\mathbf{P}[b_A(t+\Delta t) | s, {\cal B}_A(t)]$. Since $b_{A}(t)$ $=$ $(b_1(t),b_2(t))$, this probability is in fact:
\begin{align}
\mathbf{P}[b_1(t+\Delta t),b_2(t+\Delta t) | s, {\cal B}_1(t),{\cal B}_2(t)]\label{12}
\end{align} 

After some very lengthy derivation, it can be shown that the solution to the Bayesian filtering problem has the form: 
\begin{align}
 &  \mathbf{P}[b_A(t+\Delta t) | s, {\cal B}_A(t)]  =  \delta(b_1(t+\Delta t) = b_1(t) - 1,b_2(t+\Delta t) = b_2(t)+ 1)  Q_{1,A}  \nonumber\\& + \delta(b_1(t+\Delta t) = b_1(t)+1,b_2(t+\Delta t) = b_2(t) )  Q_{2,A} +  \delta(b_1(t+\Delta t) = b_1(t) - 1,\nonumber  \\  
    	&  b_2(t+\Delta t)  = b_2(t)) Q_{3,A}  + \delta(b_1(t+\Delta t) = b_1(t)+1, b_2(t+\Delta t) = b_2(t) - 1)  Q_{4,A} \nonumber  \\  
  	  	&  + \delta(b_1(t+\Delta t) = b_1(t),b_2(t+\Delta t) = b_2(t)) Q_{5,A}
 	   \label{eqn:predictbx1s}
\end{align}
The first four terms are due to the four different ways that $(b_1(t),b_2(t))$ can change according to Reactions \eqref{eq:mc1:r1} and \eqref{eq:mc1:r2}. For example, the forward reaction in \eqref{eq:mc1:r2} causes $b_2(t)$ to increase by 1 and decrease  $b_1(t)$ by 1. This is reflected in the first term. The second to fourth terms can be explained similarly.

 The expressions for $Q_{i,A}$ ($i=$1 to 5) are: 
\begin{align}
& Q_{1,A} = \lambda_2 b_1(t)  E[n_R(t) | s, {\cal B}_A(t)] \Delta t\label{eqn:q1A} \\
& Q_{2,A} = \lambda_1  (M- b_1(t)-b_2(t)) E[n_R(t) | s, {\cal B}_A(t) ] \Delta t , Q_{3,A} =  \mu_1 b_1(t) \Delta t , Q_{4,A} =  \mu_2 b_2(t) \Delta t \nonumber\\
& Q_{5,A} =  1 - ( Q_{1,A} +  Q_{2,A} + Q_{3,A} + Q_{4,A}) 
\end{align} 
As we have mentioned earlier, the derivation of $\mathbf{P}[b_A(t+\Delta t) | s, {\cal B}_A(t)]$ is lengthy. Instead of lengthy derivation, we will show in Section \ref{133} how the same result can be obtained via a graphical method. 

As explained in Section \ref{sec:MAP}, the solution to the Bayesian filtering problem is only an intermediate step of deriving the optimal demodulator for RSK modulation. The derivation of the optimal demodulation filter will be explained next.

\subsection{Demodulation filters}
\label{2222}
Once the solution to the Bayesian filtering problem has been obtained, we can put the result in \eqref{eqn:logpp_prelim} to obtain the demodulation filter for symbol $s$ by letting $\Delta t \to 0$, as follows: 
\begin{align}
&\frac{dL_s(t)}{dt} = \lim_{\Delta t \to 0} \frac {\log((\mathbf{P}[b_m(t+\Delta t) | s, {\cal B}_m(t)])}{\Delta t} + L'(t)
\label{221}
\end{align}
where $L'(t)$ is a term independent of symbol $s$. Since $L_s(t)$ does not appear on the RHS of \eqref{221} and $L'(t)$ adds the same contribution to all $L_s(t)$ for all $s = 0,...,K -1$, we can therefore ignore $L'(t)$ for the purpose of demodulation since it is the relative (rather than the absolute) magnitude of $L_s(t)$ is needed for modulation. By dropping $L'(t)$ and naming the quantity on the LHS $Z_s(t)$, we have: 
\begin{align}
&\frac{dZ_s(t)}{dt} = \lim_{\Delta t \to 0} \frac {\log(\mathbf{P}[b_m(t+\Delta t) | s, {\cal B}_m(t)])}{\Delta t} 
\label{222}
\end{align}  
For the case where $m = 1$, we put the expression of $\mathbf{P}[b_m(t+\Delta t) | s, {\cal B}_m(t)]$ in \eqref{eqn:predictb} and \eqref{eqn:predictbx}  into \eqref{222}, and we have: 
\begin{align}
&\frac{dZ_s(t)}{dt} = \frac{dD_1(t)}{dt} \log(\mu_1 b_1(t) + \lambda_{2}b_1(t)\Gamma_s(t)) +\frac{dU_1(t)}{dt} \log(\lambda_1 (M-b_1(t))\Gamma_s(t)-\lambda_1 \alpha_s(t)+
 \nonumber\\
 & \mu_{2}\beta_s(t))     -\lambda_{2}b_1(t)\Gamma_s(t) -\lambda_1 (M-b_1(t)) \Gamma_s(t) + \lambda_1 \alpha_s(t) - \mu_{2} \beta_s(t)
\label{eqn:logmap_s1}
\end{align} 
where
\begin{align}
 \Gamma_s(t) =& E[n_R(t) | s, { B_1}(t)] ,
  \beta_s(t) = E[ b_{2}(t) | s, { B_1}(t)] ,
   \alpha_s(t) =E[n_R(t) b_{2}(t) | s, { B_1}(t)] \label{eqn:opt-filter-term3} 
\end{align}

\begin{figure}
\begin{center}
\includegraphics[page=4,trim=0cm 0cm 0cm 0cm ,clip=true, width=0.5\columnwidth]{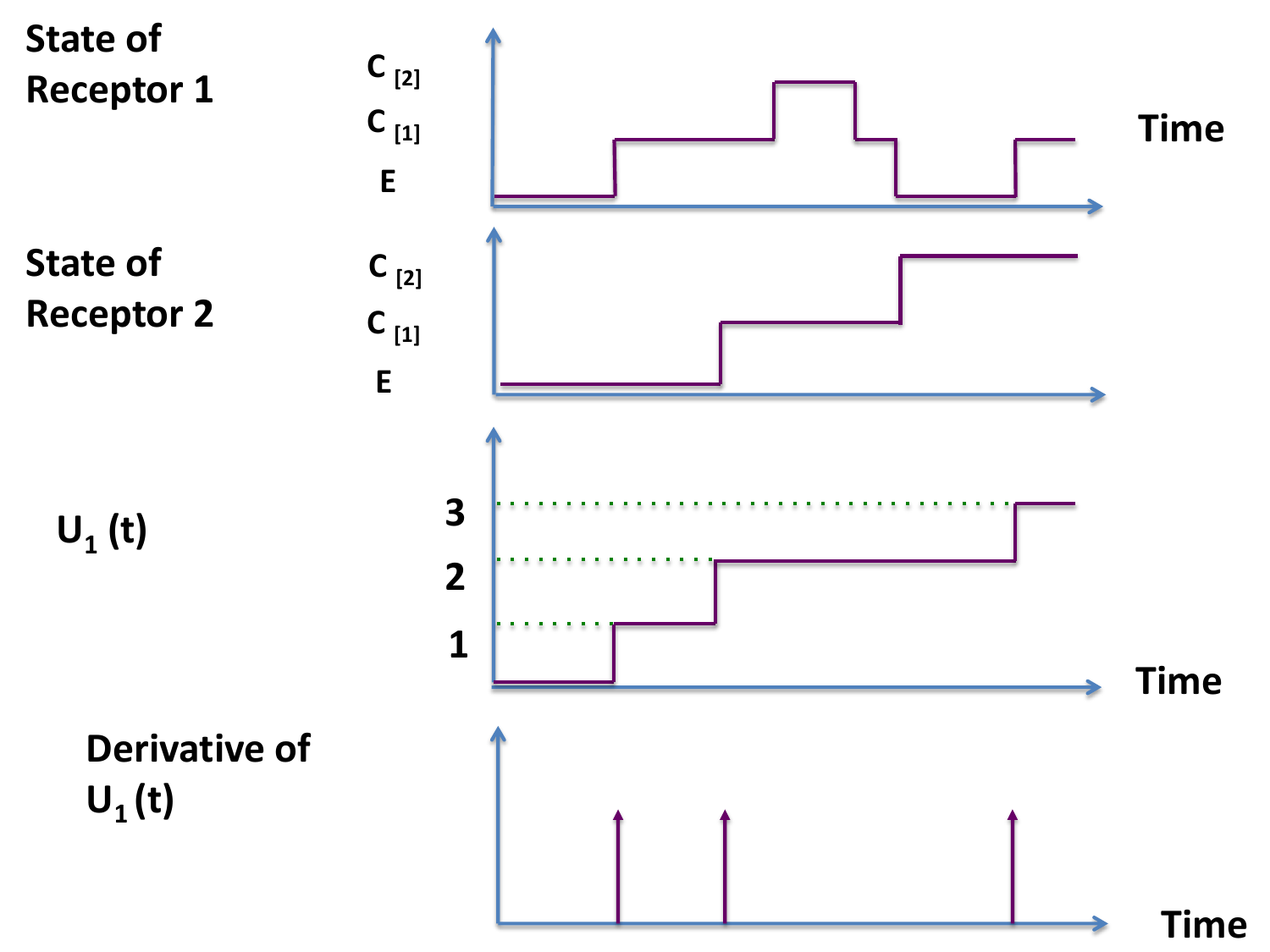}
\caption{This figure explains the meaning of $U_1(t)$, which is the cumulative
number of times that  the receptor changes from unbound state $E$ to complex $C_{[1]}$. In other words, every time when a receptor binds to form $C_{[1]}$, $U_1(t)$ is increased by 1.}
\label{fig:ut}
\end{center}
\end{figure}

In filter \eqref{eqn:logmap_s1}, $U_1(t)$ is the cumulative number of times a receptor switches from unbound state $E$ to complex $ C_{[1]}$. Fig.~\ref{fig:ut} illustrates the meaning of $U_1(t)$ when there are 2 receptors. Similarly, $D_1(t)$ is the cumulative number of times a receptor switches from  complex $C_{[1]}$ to the unbound state $E$. 

\begin{figure}
\begin{center}
\includegraphics[page=3,trim=0cm 0cm 0cm 0cm ,clip=true, width=0.45\columnwidth]{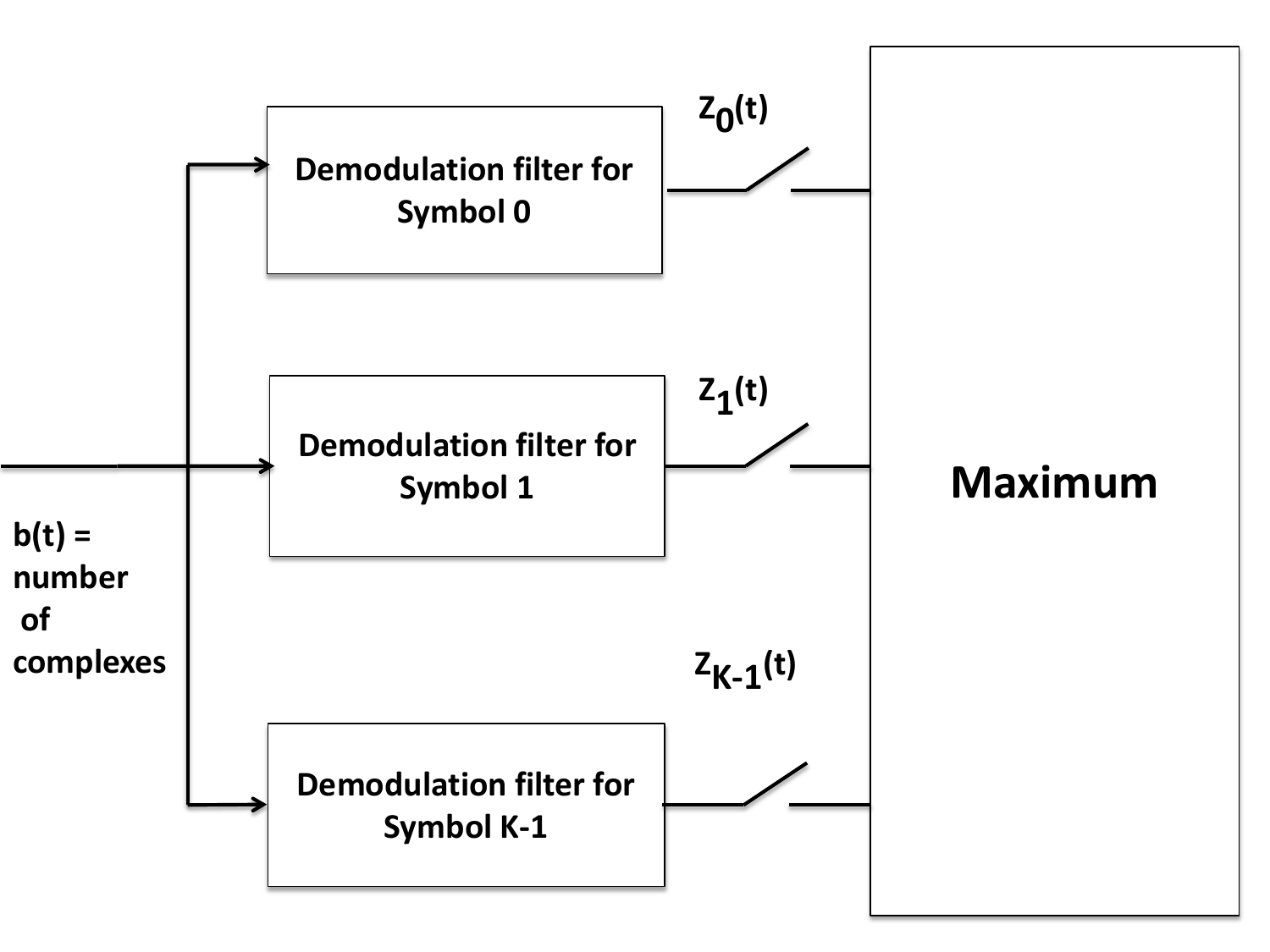}
\caption{The demodulator structure.}
\label{fig:demod}
\end{center}
\end{figure}

We initialise $Z_s(0)$ to the logarithm of the prior probability that symbol $s$ is transmitted. At time $t$, the demodulator decides that symbol $\hat{s}$ has been sent by the transmitter if:
\begin{align}
\hat{s} = {\arg\max}_{s = 0, ..., K-1} Z_s(t) 
\end{align}
Fig.~\ref{fig:demod} shows the architecture of the demodulator which runs  $K$ parallel continuous-time filters for the $K$ possible transmission symbols. The output of the demodulation filter $Z_s(t)$ is that exp$(Z_s(t))$ is proportional to the posteriori probability $P[s|B_m(t)]$.

The demodulation filters for $m = 2$ an $m = A$ can be derived similarly. Since the focus of this paper is on describing the graphical method on obtaining the solution of the Bayesian filtering problem, we will not derive these filters here. 

\subsubsection{Sub-optimal demodulation filters}
\label{sec:suboptimal}
The demodulation filters derived earlier are the optimal demodulation filters. However, the complexity of these optimal filters is high. The complexity lies with the terms  $\Gamma_s(t)$, $\beta_s(t)$ and $\alpha_s(t)$ in \eqref{eqn:opt-filter-term3}. For example, $\Gamma_s(t)$ is defined as $E[n_R(t) | s, {B_1}(t)]$, whose computation requires the solution to a Bayesian filtering problem. In our earlier work \cite{7208820}, we proposed to approximate $E[n_R(t) | s, {B_1}(t)]$ by $E[n_R(t) | s]$. Since the term $E[n_R(t) | s]$ is independent of the past history of the measurements, it does not require a filtering problem to be solved. One can view the term $E[n_R(t) | s]$ as the prior knowledge one uses in matched filters. We can use the same method to approximate $\beta_s(t)$ and $\alpha_s(t)$. In the numerical evaluation in Section \ref{sec:eval}, we will make use of the sub-optimal demodulation filters.

\subsection{Summary} This section gives an overview of how the optimal demodulator for RSK can be derived assuming that the receiver molecular circuit is a receptor with two binding sites. The key step in the derivation is to solve a Bayesian filtering problem which requires lengthy calculations. The derivation becomes even more lengthy if a molecular circuit with more chemical species and chemical reactions is used. In order to by-pass this difficulty, we present a method which allows us to write down the solution of the Bayesian filtering problem directly.

\section{General solution to the Bayesian filtering problem}
\label{133}  
We know from Section \ref{sec:model} that a key step in deriving the optimal modulation filters is to solve a Bayesian filtering problem. We have also learnt that the solution to this filtering problem requires long mathematical derivation. In this section, we present an algorithm to directly write down the solution to the Bayesian filtering problem. This algorithm can be applied to any receiver molecular circuit with any choice of measurements.

This section is structured as follows. Section \ref{qwe} presents a graph to represent the chemical reactions within the receiver molecular circuit. This graph will be used as an aid to the general solution which will be described in Section \ref{sec:graph:alg}. 
We then apply this algorithm to the ligand-receptor receiver circuit that we studied in Section \ref{sec:model}. We will show how this algorithm can be applied to all the three choices of observations that we considered earlier. 

\subsection{Chemical Reaction Graph}
\label{qwe}
In this section we present a graph to represent the chemical reactions inside the receiver molecular circuit. We will use the ligand-receptor binding with two binding sites circuit in Section \ref{sec:model} as an example. It is straightforward to generalise this to general molecular circuits. 

\begin{figure}
\begin{center}
\includegraphics[trim=0cm 0cm 0cm 0cm ,clip=true, width=0.45\columnwidth]{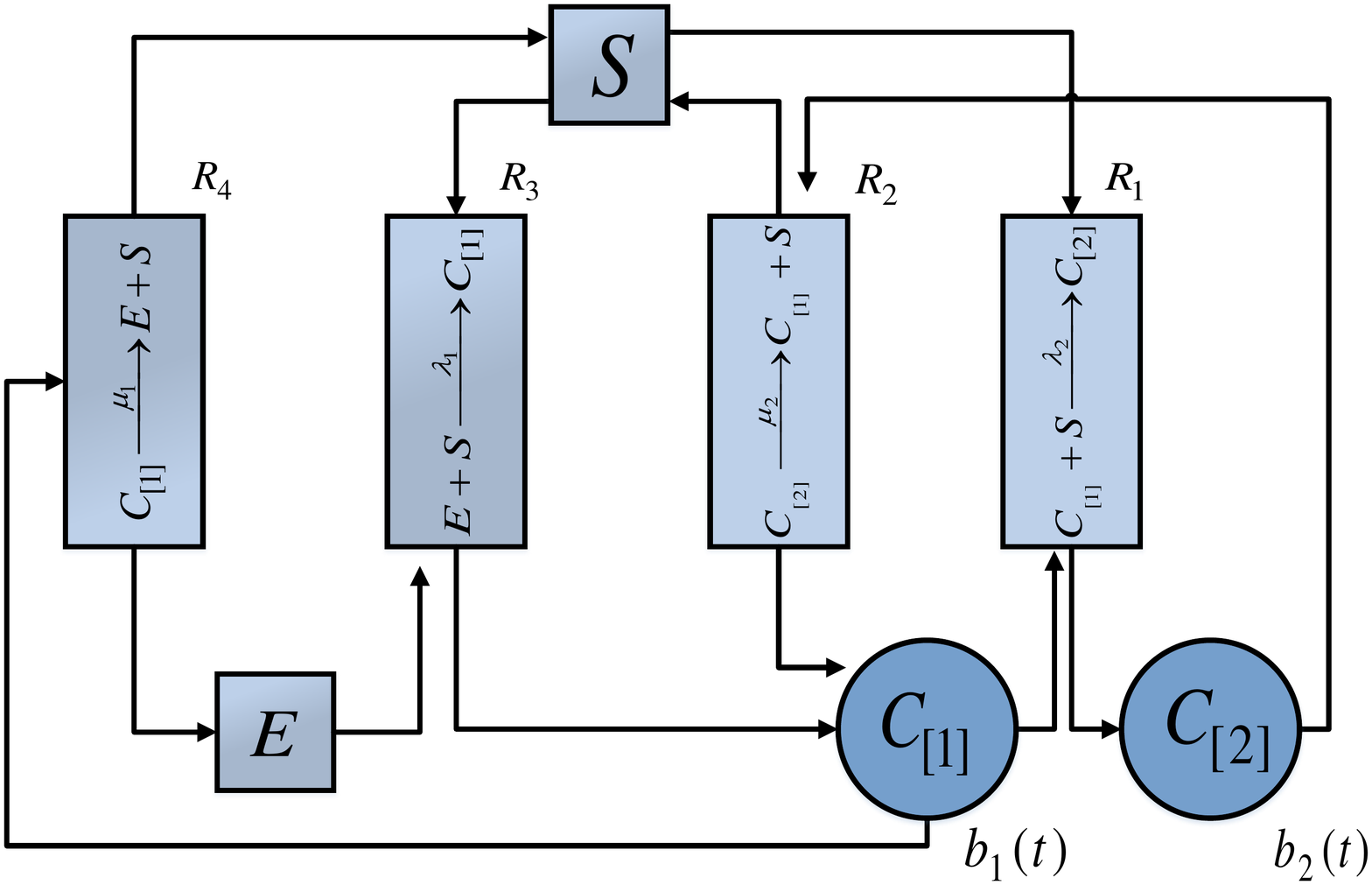}
\caption{Reaction graph when measuring both $C_{[1]}$ and $C_{[2]}$.}
\label{graph}
\end{center}
\end{figure}

Fig.~\ref{graph} shows the graph representing the ligand-receptor binding with two binding sites assuming both complexes $C_{[1]}$ and $C_{[2]}$ are measured. The graph consists of three types of nodes. The first type of nodes represent the measured chemical species found in the molecular circuit. 

These chemical species are represented by circular nodes in the figure. The second type of nodes represent unmeasured chemical species in the circuit, which in this example are signalling molecule $S$ and unbound receptor $E$. These chemical species are represented by square nodes in the figure.
The third type of nodes represent chemical reactions in the molecular circuits. There are four reactions: forward and reverse reactions in \eqref{eq:mc1:r1} and \eqref{eq:mc1:r2}. In the graph, each of these reactions is represented by a rectangular shaped node. We label the four chemical reactions by $R_1$ to $R_4$. 

The links in the reaction graph are directed. We draw a link from a reaction node (a rectangular node) to a chemical species node (a circular or square node) if the chemical species is a product of that chemical reaction. Similarly, we draw a link from a chemical species node to a reaction node if the chemical species is a reactant of that chemical reaction. In Fig.~\ref{graph}, we use reaction $R_1$, which is the right-most rectangular box, as an example. The reaction is $C_{[1]} + S \rightarrow C_{[2]}$. We see that there is a directed link from the $R_1$ reaction node to the chemical species node $C_{[2]}$ which is the product of this reaction. We can also see directed links from the chemical species nodes $S$ and $C_{[1]}$ to the $R_1$ reaction node because they are the reactants of this reaction. 

\subsection{General Algorithm} 
\label{sec:graph:alg} 
In this section, we describe a general algorithm to write down the solution to the Bayesian filtering problem that we need to solve in order to obtain the demodulation filter. 
\textcolor{black}{The proof of the general algorithm can be found in Appendix A}. 

We begin by defining a number of notation. 
We assume the demodulator observes (or measures) $m_O$ chemical species in the receiver molecular circuit. We will denote these $m_O$ observed chemical species as $O_1$, $O_2$, ... , $O_{m_O}$. These $m_O$ observed species are involved in $m_R$ chemical reactions in the molecular circuit. Note that $m_R$ is not necessary the same as the number of chemical reactions in the molecular circuits because some chemical reactions may not involve any observed species. We will denote these $m_R$ reactions by $R_1$, $R_2$, ..., $R_{m_R}$. We assume that there are altogether $m_U$ unobserved species that are involved in these $m_R$ reactions. We will denote these $m_U$ unobserved species by $U_1$, $U_2$, ... , $U_{m_U}$. 

We assume that Reaction $R_i$ has the general form: 
\begin{align}
\cee{
& a_{i,1} O_1 + ... + a_{i,m_O} O_{m_O} + b_{i,1} U_1 + ... + b_{i,m_U} U_{m_U}   ->[\kappa_i]  \nonumber\\
 &  c_{i,1} O_1 + ... + c_{i,m_O} O_{m_O} + d_{i,1} U_1 + ... + d_{i,m_U} U_{m_U}
\label{eq:gen_reaction}
} 
\end{align}
where $a_{i,j}$, $b_{i,j}$, $c_{i,j}$ and $d_{i,j}$ are non-negative integers. In Reaction $R_i$, we have $a_{i,1}$ molecules of $O_1$, ..., $a_{i,m_O}$ molecules of $O_{m_O}$, $b_{i,1}$ molecules of $U_1$, ... , $b_{i,m_U}$ molecules of $U_{m_U}$ reacting to produce $c_{i,1}$ molecules of $O_1$, ..., $c_{i,m_O}$ molecules of $O_{m_O}$, $d_{i,1}$ molecules of $U_1$, ... , $d_{i,m_U}$ molecules of $U_{m_U}$. Note that the coefficients $a_{i,j}$, $b_{i,j}$, $c_{i,j}$ and $d_{i,j}$ can be zero. A zero $a_{i,j}$ or $b_{i,j}$ (resp. $c_{i,j}$ or $d_{i,j}$) coefficient means that the chemical species is {\bf not} consumed (produced) in the reaction. E.g., if $a_{i,1}$ is zero, then $O_1$ is not a reactant in Reaction $R_i$. Since chemical reactions generally involve only one or two reactants, and also a small number of products, we expect most of the $a_{i,j}$, $b_{i,j}$, $c_{i,j}$ and $d_{i,j}$ are zero. Although the general form of chemical reaction in \eqref{eq:gen_reaction} appear to use many chemicals, the reality is that only a small number of chemicals are involved in each reaction. The dense form of \eqref{eq:gen_reaction} may not be the best way to represent chemical reactions, so we will make use of the reaction graph describe in Section \ref{qwe} to aid the explanation.

The Bayesian filtering problem to be solved is to determine the joint probability of the counts of the chemical species  $O_1$, $O_2$, ... , $O_{m_O}$ at time $t + \Delta t$ given the transmission symbol and the history of the counts of chemical species  $O_1$, $O_2$, ... , $O_{m_O}$ up till time $t$. We will denote this probability by $P$ and it consists of the sum of $m_R+1$ terms. The first $m_R$ terms come from the $m_R$ reactions while the $(m_R+1)$-th term comes from no reactions taking place in the time interval $(t, t + \Delta t)$. Let $n_{O_j}(t)$ and $n_{U_j}(t)$ denote, respectively, the number of $O_j$ and $U_j$ molecules at time $t$. 

We now explain how the $i$-th term in $P$ is obtained with $1 \leq i \leq m_R$. The $i$-th term in $P$ comes from Reaction $R_i$. If Reaction $R_i$ happens, the net change in the number of $O_j$ molecules is $c_{i,j} - a_{i,j}$. Every term in $P$ is a product of an indicator function $\delta(...)$ and $Q_{i,m}$. (Note: The subscript $m$ is used to denote dependence on the choice of measurement, which is also used in Section \ref{24}.)  The term $Q_{i,m}$ depends on reaction rate and will be explained a moment. 
The indicator function is:
\begin{align}
\delta( & n_{O_1}(t + \Delta t) = n_{O_1}(t) + c_{i,1} - a_{i,1}, ..., n_{O_{m_O}}(t + \Delta t) = n_{O_{m_O}}(t) + c_{i,m_O} - a_{i,m_O}   )
\end{align} 
which takes the value of 1 if Reaction $R_i$ has occurred. Note that the inputs to the indicator function contain only the counts of the measured species. Given that chemical reactions generally have only up to two reactants per reaction, most of the $(c_{i,j} - a_{i,j})$ are zero. A more efficient method to identify the non-zero $(c_{i,j} - a_{i,j})$ is to use a graph. This can be done by first identifying the reaction nodes in the graph, and then find out the chemical species that participate in this reaction by using the links that are connected to the reaction nodes. 

By the Law of Mass Action, the reaction rate of Reaction $R_i$ at time $t$ is:
\begin{align}
  \kappa_i  \Pi_{j = 1}^{m_O}  n_{O_j}(t)^{a_{i,j}}   \Pi_{j = 1}^{m_U}  n_{U_j}(t)^{b_{i,j}}
  \label{qw}
\end{align}
Note that this is the general form of reaction rate based on the general reaction in \eqref{eq:gen_reaction}.  Recall that many $a_{i,j}$ and $b_{i,j}$ are zero, therefore the above reaction rate depends only on the counts of a small number of chemical species. Since only the reactants are needed to calculate the reaction rate, one can easily identify the reactants from the reaction graph by tracing the incoming links to the reaction nodes.  

We also need the mean reaction rate of $R_i$ conditioned on the symbol $s$ that transmitter has sent and the past history of measurement (denoted by ${\cal B}(t)$). We denote the product of $\Delta t$ and this conditional mean by the term $Q_{i,m}$. The reader will see that this $Q_{i,m}$ will have a one-to-one correspondence with the $Q_{i,m}$ terms in Section \ref{24}, hence the same notation is used. The expression for general $Q_{i,m}$  is: 
\begin{align}
   & E[  \kappa_i   \Pi_{j = 1}^{m_O}  n_{O_j}(t)^{a_{i,j}}   \Pi_{j = 1}^{m_U}  n_{U_j}(t)^{b_{i,j}} | s, {\cal B}(t)] \Delta t =  \kappa_i   \Pi_{j = 1}^{m_O}  n_{O_j}(t)^{a_{i,j}}   E[\Pi_{j = 1}^{m_U}  n_{U_j}(t)^{b_{i,j}}| s, {\cal B}(t)] \Delta t
   \label{29}
\end{align}
Note that the equality holds because the chemical species $O_j$ are measured, so their counts form part of the history of measurement. This completes the description of the first $m_R$ terms in $P$. 

The $(m_R+1)$-th term of $P$ is the product of an indicator function and $Q_n$ = $[1 - (\sum_{i = 1}^{m_R} {Q_{i,m}}) ]$. This term corresponds to the case that no chemical reaction has taken place. The indicator function is:
\begin{align}
\delta( n_{O_1}(t + \Delta t) = n_{O_1}(t) , ...,   
           n_{O_{m_O}}(t + \Delta t) = n_{O_{m_O}}(t))
\end{align} 

As mentioned before, $P$ is the sum of the $(m_R+1)$ terms as described above. 
\begin{figure}
\begin{center}
\includegraphics[trim=0cm 0cm 0cm 0cm ,clip=true, width=0.45\columnwidth]{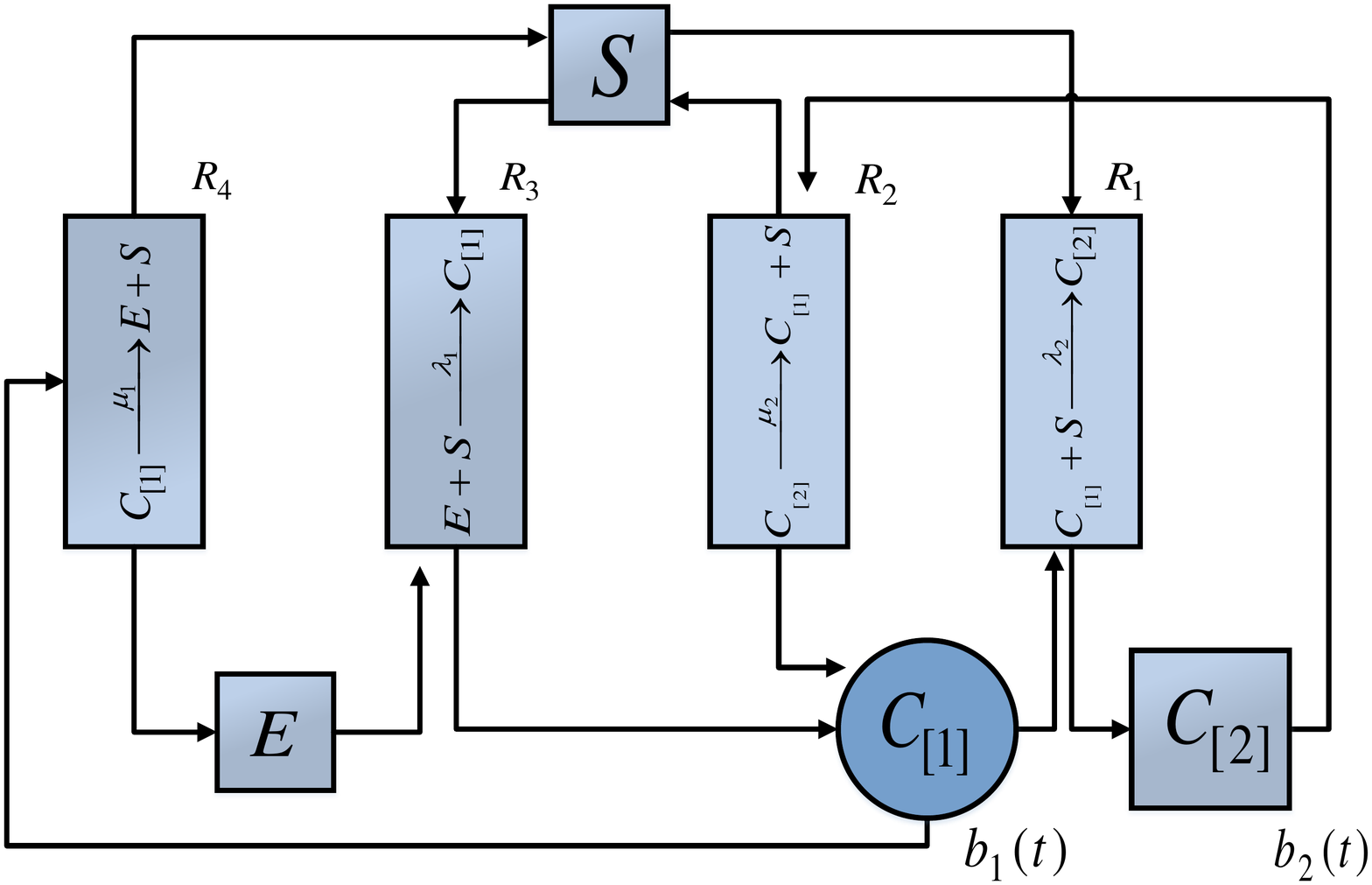}
\caption{Reaction graph when measuring $C_{[1]}$ only.}
\label{fig:graph1}
\end{center}
\end{figure}
\subsection{Applying the general algorithm} 
\label{33}
In this section, we apply the general algorithm that we described earlier to the ligand-receptor binding with two binding site circuit that we consider in Section \ref{sec:model}. We have already presented the solution to the Bayesian filtering problem for this molecular circuit earlier and the reader should be able to verify that the graphical model gives the same answer. 

We will divide the example into three cases according to the three choices of measurements: (a) Measuring the number of complexes $C_{[1]}$ only;  (b) Measuring the number of complexes $C_{[2]}$ only; (c) Measuring both the number of complexes $C_{[1]}$ and $C_{[2]}$.

\begin{table}[]
\centering

\begin{tabular}{|c|c|c|c|c|c|c|c|c|c|c}
\hline
\multicolumn{2}{|c|}{}	&	\multicolumn{1}{|c|}{$O_1$ $\leftrightarrow$ $C_{[1]}$}		&	\multicolumn{1}{|c|}{$U_1$ $\leftrightarrow$ $S$}		 &	\multicolumn{1}{|c|}{$U_2$ $\leftrightarrow$ $E$ } &	\multicolumn{1}{|c|}{$U_3$ $\leftrightarrow$ $C_{[2]}$} & \multicolumn{1}{|c|}{$O_1$ $\leftrightarrow$ $C_{[1]}$}		&	\multicolumn{1}{|c|}{$U_1$ $\leftrightarrow$ $S$}		 &	\multicolumn{1}{|c|}{$U_2$ $\leftrightarrow$ $E$ } &	\multicolumn{1}{|c|}{$U_3$ $\leftrightarrow$ $C_{[2]}$}	\\
\hline
$R_i$	&	$\kappa_i$		& $a_{i,1}$  & $b_{i,1}$ & $b_{i,2}$ & $b_{i,3}$ & $c_{i,1}$  & $d_{i,1}$ & $d_{i,2}$  & $d_{i,3}$		\\
\hline
$R_{1}$      & $\lambda_2$ & 1     & 1     & 0     & 0     & 0     & 0    & 0    & 1    \\ \hline
$R_{2}$      & $\mu_2$     & 0     & 0     & 0     & 1     & 1     & 1    & 0    & 0    \\ \hline
$R_{3} $     & $\lambda_1$ & 0     & 1     & 1     & 0     & 1     & 0    & 0    & 0    \\ \hline
$R_{4} $     & $\mu_1$     & 1     & 0     & 0     & 0     & 0     & 1    & 1    & 0    \\ \hline
\end{tabular}
\caption{Measuring $C_{[1]}$ only}
\label{tab:C1only}
\end{table}

\subsubsection{Measuring the number of complexes $C_{[1]}$ only}
Since only $C_{[1]}$ is measured, there is $m_O = 1$ measured species. All the four chemical reactions in the molecular circuit involve $C_{[1]}$, hence $m_R = 4$. These four reactions involve $m_U = 3$ unmeasured species, which are $S$, $E$ and $C_{[2]}$. In the general algorithm, we have used $O_j$ and $U_j$ to label, respectively, the measured and unmeasured species. We will identify $O_1$ as $C_{[1]}$, as well as $U_1$, $U_2$ and $U_3$, respectively, as $S$, $E$ and $C_{[2]}$. The mapping between general species names used in the general algorithm and the actual chemical species is shown in Table \ref{tab:C1only}. 

The graph that corresponds to this choice of measurement is shown in Fig.~\ref{fig:graph1}. Note that there is one circular (measured species) node, 3 square (unmeasured species) nodes and 4 rectangular (reaction) nodes. We label the four reactions using $R_1$, $R_2$, $R_3$ and $R_4$ in Fig.~\ref{fig:graph1}. All these reactions can be written in the general form \eqref{eq:gen_reaction}. Table \ref{tab:C1only} shows the coefficients of these reactions when expressed in the general form. For example, Reaction $R_1$ is 
\begin{align}
\cee{
C_{[1]} + S ->[\lambda_2] C_{[2]}
} 
\end{align} 
and when expressed in the general form, without showing the zero coefficients for brevity, is: 
\begin{align}
\cee{
a_{1,1} O_1 + b_{1,1} U_1 ->[\kappa_1] d_{1,3} U_{3} 
} 
\end{align} 
where $a_{1,1}$, $b_{1,1}$ and $d_{1,3}$ equal to 1. 

Given that there are 4 reactions involving $C_{[1]}$, the probability $\mathbf{P}[b_1(t+\Delta t) | s, {\cal B}_1(t)]$ is a sum of 5 terms. The first term comes from Reaction $R_1$. This term consists of the product of an indicator function, a conditional mean reaction rate and $\Delta t$. The indicator function shows the change in the number of measured species when Reaction $R_1$ takes place, hence it is given by:
\begin{align}
\delta(b_1(t+\Delta t) = b_1(t) - 1) \nonumber 
\end{align} 
The rate of reaction $R_1$ is determined by the number of reactant molecules. It can be seen from the reaction graph in Fig.~\ref{fig:graph1} that the reactants of this reaction are $C_{[1]}$ and $S$. The number of $C_{[1]}$ molecules at time $t$ is $b_1(t)$ as defined earlier. The number of signalling molecules in the receiver voxel is $n_R(t)$. Hence the reaction rate is  $\lambda_2  b_1(t) n_R(t)$. Conditioning this on the transmitted symbol $s$ and the past history ${\cal B}_1(t)$, the required conditional mean reaction rate is:
\begin{align}
\lambda_2  b_1(t) E[n_R(t) | s, {\cal B}_1(t)] \nonumber 
\end{align} 
Note that the product of this conditional mean reaction rate expression with $\Delta t$ gives the same result as $Q_{1,1}$ in \eqref{eqn:q11}. Hence the term due to Reaction $R_1$ is: 
\begin{align}
\delta(b_1(t+\Delta t) = b_1(t) + 1) \times  \{ \lambda_2  b_1(t) E[n_R(t) | s, {\cal B}_1(t)]  \}  \times \Delta t \nonumber 
\end{align} 
 
We can similarly work out the next three terms corresponding to Reactions $R_2$ to $R_4$. Once we have worked out these terms, we can readily obtain the last term which corresponds to no reaction had taken place.

\begin{figure}
\begin{center}
\includegraphics[trim=0cm 0cm 0cm 0cm ,clip=true, width=0.45\columnwidth]{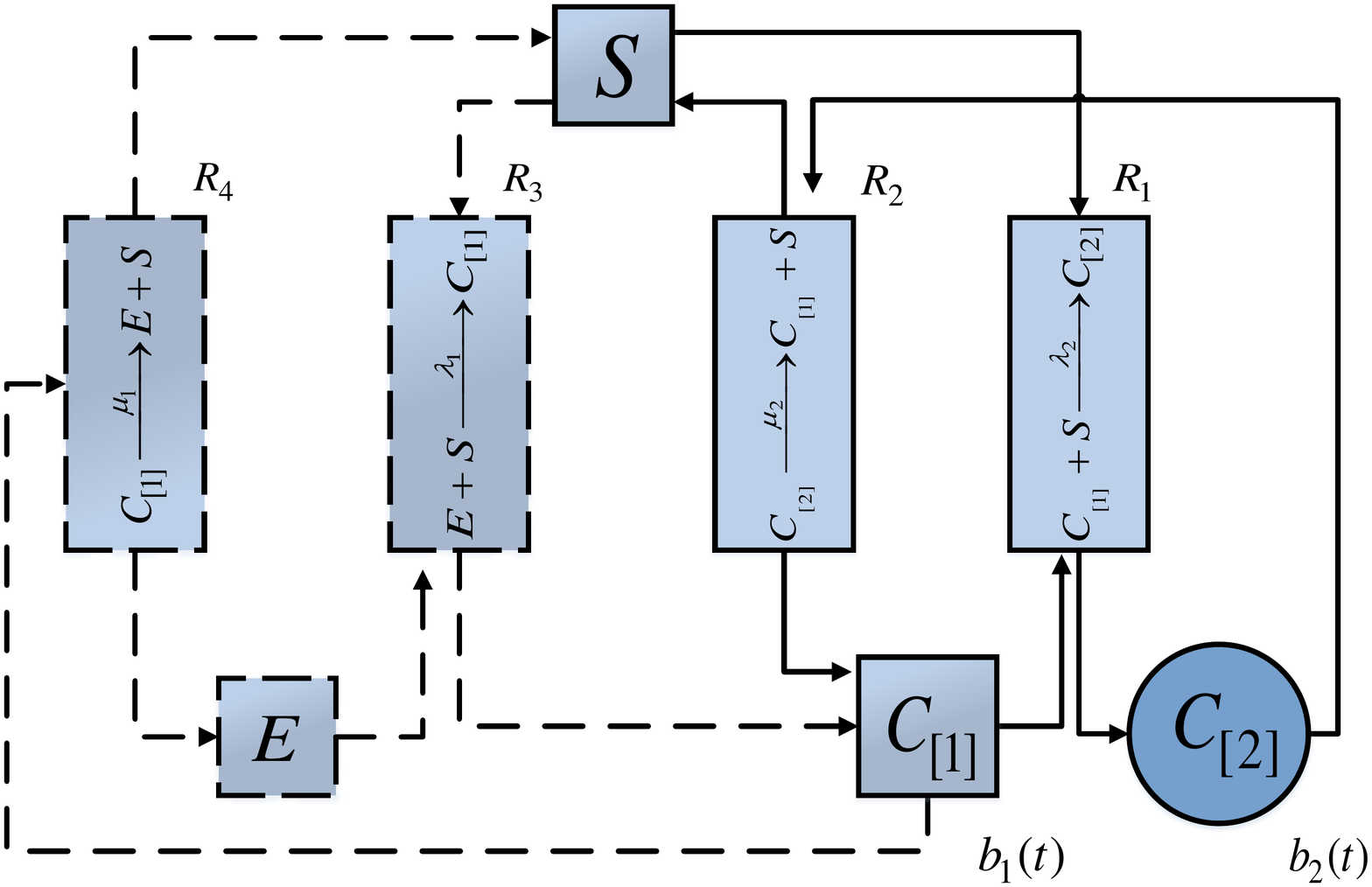}
\caption{Reaction graph when measuring $C_{[2]}$ only. Note that the dashed line nodes and links are not part of the reaction graph. They are shown in this figure to indicate what has been taken out of the complete reaction graph.}
\label{fig:graph2}
\end{center}
\end{figure}

\begin{table}[]
\centering

\begin{tabular}{|c|c|c|c|c|c|c|c|c|}
\hline
\multicolumn{2}{|c|}{}	&	\multicolumn{1}{|c|}{$O_1$ $\leftrightarrow$ $C_{[2]}$}		&	\multicolumn{1}{|c|}{$U_1$ $\leftrightarrow$ $S$}		 &	\multicolumn{1}{|c|}{$U_2$ $\leftrightarrow$ $C_{[1]}$ } &		\multicolumn{1}{|c|}{$O_1$ $\leftrightarrow$ $C_{[2]}$}		&	\multicolumn{1}{|c|}{$U_1$ $\leftrightarrow$ $S$}		 &	\multicolumn{1}{|c|}{$U_2$ $\leftrightarrow$ $C_{[1]}$ }	\\
\hline
$R_i$ & $\kappa_i$ & $a_{i,1}$  & $b_{i,1}$ & $b_{i,2}$ & $c_{i,1}$  & $d_{i,1}$ & $d_{i,2}$  
\\ \hline
$R_{1}$      & $\lambda_2$ & 0          & 1     & 1     & 1       & 0    & 0    \\ \hline
$R_{2}$      & $\mu_2$     & 1         & 0     & 0     & 0         & 1    & 1    \\ \hline
\end{tabular}
\caption{Measuring $C_{[2]}$ only}
\label{tab:C2only}
\end{table}

\subsubsection{Measuring the number of complexes $C_{[2]}$ only}
Since only $C_{[2]}$ is measured, there is $m_O = 1$ measured species. Out of the four reactions in the molecular circuit, only two of them involve the measured species $C_{[2]}$, hence $m_R = 2$. These two reactions involve $m_U = 2$ unmeasured species, which are $S$ and $C_{[2]}$. We will identify $O_1$ as $C_{[2]}$; as well as identifying $U_1$ and $U_2$, respectively, as $S$ and $C_{[1]}$. The mapping between general species names used in the general algorithm and the actual chemical species is shown in Table \ref{tab:C2only}. 

The graph that corresponds to this choice of measurement is shown in Fig.~\ref{fig:graph2}. Note that there is one circular (measured species) node, 2 square (unmeasured species) nodes and 2 rectangular (reaction) nodes. Note that the dashed line nodes and links are not part of the reaction graph. In other words, only the nodes and links in solid lines are part of the graph. We have deliberately used the dashed lines to indicate what has been taken out of the complete reaction graph. We label the two reactions using $R_1$ and $R_2$ in Fig.~\ref{fig:graph2}. All these reactions can be written in the general form \eqref{eq:gen_reaction}. Table \ref{tab:C2only} shows the coefficients of these reactions when expressed in the general form. 

Given that there are 2 reactions involving $C_{[2]}$, the probability $\mathbf{P}[b_2(t+\Delta t) | s, {\cal B}_2(t)]$ is a sum of 3 terms. The first two terms come from Reactions $R_1$ and $R_2$. The method to work out these terms is exactly the same as that used in the case when $C_{[1]}$ is measured, so we will not repeat. 

Although the cases of measuring $C_{[1]}$ only and $C_{[2]}$ only are similar, note that there are two main differences. First, when measuring $C_{[2]}$ only, we do not need to consider all the reactions in the molecular circuit but only those two that involve the measured species $C_{[2]}$. Second, when measuring $C_{[2]}$ only, we  only consider the solid lines in the graph shown in Fig.~\ref{fig:graph2}. The dashed lines represent the reactions not involved in this case.

\subsubsection{Measuring both $C_{[1]}$ and $C_{[2]}$}
Since both $C_{[1]}$ and $C_{[2]}$ are measured, $m_O = 2$. These two measured species are involved in all the four chemical reactions in the molecular circuit, hence $m_R = 4$. These four reactions involve $m_U = 2$ unmeasured species, which are $S$ and $E$. We will identify $O_1$ and $O_2$ as, respectively, $C_{[1]}$ and $C_{[2]}$; as well as $U_1$ and $U_2$, respectively, as $S$ and $E$. The mapping between general species names used in the general algorithm and the actual chemical species is shown in Table \ref{tab:C1andC2}. 
The graph that corresponds to this choice of measurement is shown in Fig.~\ref{graph}. Note that there are 2 circular (measured species) node, 2 square (unmeasured species) nodes and 4 rectangular (reaction) nodes. We label the four reactions using $R_1$, $R_2$, $R_3$ and $R_4$ in Fig.~\ref{graph}. All these reactions can be written in the general form \eqref{eq:gen_reaction}. Table \ref{tab:C1andC2} shows the coefficients of these reactions when expressed in the general form. 

Given that there are 4 reactions involving $C_{[1]}$ and $C_{[2]}$, the probability $\mathbf{P}[b_A(t+\Delta t) | s, {\cal B}_A(t)]$ is a sum of 5 terms. The first term comes from Reaction $R_1$, which is:
\begin{align}
\cee{
C_{[1]} + S ->[\lambda_2] C_{[2]}
} 
\end{align} 
This term consists of the product of an indicator function, a conditional mean reaction rate and $\Delta t$. The indicator function shows the change in the number of measured species $C_{[1]}$ and $C_{[2]}$ when Reaction $R_1$ takes place, hence it is given by:
\begin{align}
\delta(b_1(t+\Delta t) = b_1(t) - 1, b_2(t+\Delta t) = b_2(t) + 1) \nonumber 
\end{align} 
because one $C_{[1]}$ molecule is consumed and one $C_{[2]}$ molecule is produced in Reaction $R_1$. It can be seen from the reaction graph in Fig.~\ref{graph} that the reactants of this reaction are $C_{[1]}$ and $S$. Hence the reaction rate is  $\lambda_2  b_1(t) n_R(t)$. Conditioning this on the transmitted symbol $s$ and the past history ${\cal B}_A(t)$, the required conditional mean reaction rate is:
\begin{align}
\lambda_2  b_1(t) E[n_R(t) | s, {\cal B}_A(t)] \nonumber 
\end{align} 
Note that the product of this conditional reaction rate with $\Delta t$ gives the same result as $Q_{1,A}$ in \eqref{eqn:q1A}. Hence the term due to Reaction $R_1$ is:
\begin{align}
& \delta(b_1(t+\Delta t) = b_1(t) - 1, b_2(t+\Delta t) = b_2(t) + 1) \times   
 \{ \lambda_2  b_1(t) E[n_R(t) | s, {\cal B}_A(t)]  \}  \times \Delta t \nonumber 
\end{align} 
 
 \begin{table}[]
 \centering
 
 \begin{tabular}{|c|c|c|c|c|c|c|c|c|c|c}
 \hline
 \multicolumn{2}{|c|}{}	&	\multicolumn{1}{|c|}{$O_1$ $\leftrightarrow$ $C_{[1]}$}		&		\multicolumn{1}{|c|}{$O_2$ $\leftrightarrow$ $C_{[2]}$} & \multicolumn{1}{|c|}{$U_1$ $\leftrightarrow$ $S$}		 &	\multicolumn{1}{|c|}{$U_2$ $\leftrightarrow$ $E$ } &\multicolumn{1}{|c|}{$O_1$ $\leftrightarrow$ $C_{[1]}$}		&	\multicolumn{1}{|c|}{$O_2$ $\leftrightarrow$ $C_{[2]}$} & \multicolumn{1}{|c|}{$U_1$ $\leftrightarrow$ $S$}		 &	\multicolumn{1}{|c|}{$U_2$ $\leftrightarrow$ $E$ } 		\\
 \hline
 $R_i$ & $\kappa_i$& $a_{i,1}$ & $a_{i,2}$ & $b_{i,1}$ & $b_{i,2}$ & $c_{i,1}$ & $c_{i,2}$ & $d_{i,1}$ & $d_{i,2}$  
 \\ \hline
 $R_{1}$      & $\lambda_2$ & 1     & 0     & 1    & 0     & 0     & 1    & 0    & 0    \\ \hline
 $R_{2}$      & $\mu_2$     & 0     & 1     & 0     & 0     & 1     & 0    & 1    & 0    \\ \hline
 $R_{3} $     & $\lambda_1$ & 0     & 0     & 1     & 1     & 1     & 0    & 0    & 0    \\ \hline
 $R_{4} $     & $\mu_1$     & 1     & 0     & 0     & 0     & 0     & 0    & 1    & 1    \\ \hline
 \end{tabular}
 \caption{Measuring both $C_{[1]}$ and $C_{[2]}$}
 \label{tab:C1andC2}
 \end{table}

We can similarly work out the next three terms corresponding to Reactions $R_2$ to $R_4$. Once we have worked out these terms, we can readily obtain the last term which corresponds to no reaction had taken place. 
Note that this example, which has two measured species, shows how our general algorithm can be used when multiple species are measured.

\section{Numerical Examples}
\label{sec:eval}
The previous section has presented a general algorithm to solve the Bayesian filtering problem required for deriving the demodulation filters. We illustrated the algorithm by applying it to a ligand-receptor binding circuit with two binding sites. Since the algorithm can be applied to any molecular circuit and any choice of measurements, in this section we will apply the algorithm to two new molecular circuits. Both molecular circuits are ligand-receptor binding, one with three binding sites and another five. We will study the impact of the choice of measurements on the symbol error rate (SER). We first describe the methodology and follow by the results. 

\subsection{Methodology} 

We assume the medium has a size of 2$\mu$m $\times$ 2$\mu$m $\times$ 1 $\mu$m. The voxel size is ($\frac{1}{3}$$\mu$m)$^{3}$ (i.e. $W = \frac{1}{3}$ $\mu$m). This forms a grid of $6 \times 6 \times 3$ voxels. We assume the transmitter and receiver are located at voxels (2,3,2) and (5,3,2) respectively according to their positions in the grid of voxels. The diffusion coefficient of the propagation medium is 1 $\mu$m$^2$s$^{-1}$. 
 
We assume an absorbing boundary condition where ligands may leave the surface of boundary voxel at a rate $\frac{d}{50}$. The transmitter is assumed to use $K = 2$ symbols. Each symbol is represented by an emission pattern which is generated by a chemical reaction of the form \eqref{eq:t1:1}. \textcolor{black}{Both Symbols use
this reaction such that Symbols 0 and 1 causes, respectively, 10 and 20 signalling molecules to be generated per second on average by the transmitter. The parameters for the receiver molecular circuits, i.e. the kinetic parameters for the reactions and the number of receptors, will vary from experiment to experiment. These parameters are specified in later sections.}
 
We use Stochastic Simulation Algorithm (SSA) \cite{Gillespie:1977ww} to simulate the CTMP that models both diffusion and reaction of molecules in the system. In order to use the sub-optimal demodulation filter described in Section \ref{sec:suboptimal}, we require the mean of a few quantities. For example, the sub-optimal filter in Section \ref{2222} requires $E[n_R(t) | s]$, $E[ b_{2}(t) | s]$ and $E[n_R(t) b_{2}(t) | s]$. (Note that these quantities are obtained by dropping ${B_1}(t)$ from the conditioning part of these expectations in 
\eqref{eqn:opt-filter-term3}.) Unfortunately, it is not possible to analytically compute such mean quantities from the CTMP because of moment closure problem which arises when the transition rate is a non-linear function of the state \cite{Smadbeck:2013fk}. We therefore resort to simulation to estimate these mean quantities by running SSA simulation 500 times and use the results to estimate the mean. Note that these simulations are different from those used for performance evaluation. 

We numerically integrate the sub-optimal demodulation filters to obtain $Z_s(t)$. We use the initial condition $Z_s(0) = 0$ for all $s$ which means that all symbols are equally probable in the system. We will use SER as the performance metric. 

\subsection{Ligand-receptor binding with 3 binding sites}
We study the performance of the sub-optimal demodulator assuming that the front-end molecular circuit consists of ligand-binding receptors with 3 binding sites. Comparing to the ligand-binding receptor with 2 binding site example earlier, this circuit has one additional chemical species $C_{[3]}$ which is formed by binding three signalling molecules to the receptor. This circuit has an additional chemical reaction, in addition to Reactions \eqref{eq:mc1:r1} and \eqref{eq:mc1:r2}, which is:
\begin{align}
\cee{
S + C_{[2]} &<=>C[\tilde{\lambda}_3][\mu_3] C_{[3]} \label{eq:mc1:r3} 
}
\end{align}
\textcolor{black}{
where $\tilde{\lambda}_3$ and $\mu_3$ are reaction rate constants. Similarly, we define $\lambda_3 = \frac{\tilde{\lambda}_3}{W^3}$.}
We consider four different choices of measurements: (a) Measuring the number of complexes $C_{[1]}$ only; (b) Measuring the number of complexes $C_{[2]}$ only; (c) Measuring the number of complexes $C_{[3]}$ only; and, (d) Measuring the number of all three complexes. In this study, we keep the reaction constants $\mu_1$, $\mu_2$ and $\mu_3$ to fixed values of $1$. We also fix the reaction constant $\lambda_1$ to a value of $1$. We vary three parameters $k_1$  = $\frac{\lambda_2}{\lambda_1}$, $k_2$ =  $\frac{\lambda_3}{\lambda_1}$ and the number of receptors $M$. 

\begin{figure}
\begin{center}
\includegraphics[trim=0cm 0cm 0cm 0cm ,clip=true, width=0.5\columnwidth]{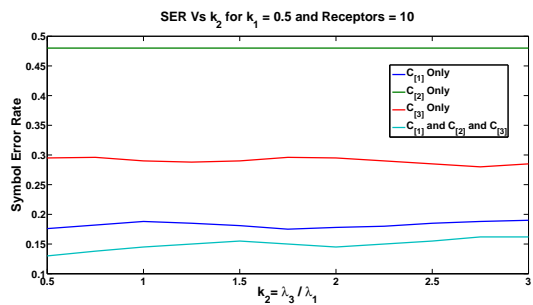}
\caption{\textcolor{black} {Impact of the choice of measurements for the 3-binding site case. Symbol Error Rate vs $k_2$ , for $k_1=0.5$ , $M=10$}}
\label{8}
\end{center}
\end{figure}

Fig~\ref{8} shows SER for the four choices of measurements for $k_1 = 0.5$, $k_2$ varying from 0.5 to 3, and $M = 10$. If the counts of all the three complexes are measured, then the SER is the lowest. This is intuitive because measuring all species gives the highest amount of information. We therefore can use this particular choice of measurements as the benchmark. When we measure the counts of only one type of complex, we find that we get the lowest SER by measuring $C_{[1]}$, followed by $C_{[3]}$ and the worst is by measuring $C_{[2]}$. We vary the values of $k_1$, $k_2$ and $M$, we find that this order of performance remains the same. Results have not been included due to space limit. 


\subsection{Ligand-receptor binding with 5 binding sites}
This section considers a molecular circuit consisting of ligand-binding receptors with 5 binding sites. Comparing to the ligand-binding receptor with 3 binding sites example earlier, this circuit has two additional chemical species $C_{[4]}$ and $C_{[5]}$, which is formed by binding of, respectively, four and five signalling molecules to the receptor. 

We consider six different choices of measurements: the first five choices are measuring exactly one of the five species $C_{[1]}$, $C_{[2]}$, $C_{[3]}$, $C_{[4]}$ and $C_{[5]}$; the sixth choice is to measure all the five species. In this study, we keep the reaction constants $\mu_1$, $\mu_2$ , $\mu_3$, $\mu_4$ and $\mu_5$ to a fixed value of $1$. We also keep the reaction constants $\lambda_1$,  $\lambda_3$, $\lambda_4$ and $\lambda_5$ to a fixed value of $1$. For this case, we vary only two parameters: the ratio $k$  = $\frac{\lambda_2}{\lambda_1}$ and the number of receptors $M$. 

\begin{figure}
\begin{center}
\includegraphics[trim=0cm 0cm 0cm 0cm ,clip=true, width=0.5\columnwidth]{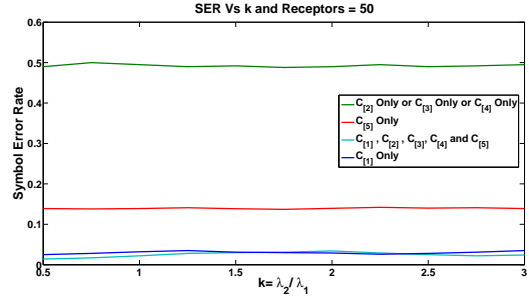}
\caption{\textcolor{black}{Impact of the choice of measurements for the 5-binding site case. Symbol Error Rate vs $k$ for  $M=50$.}}
\label{13}
\end{center}
\end{figure}

Figures \ref{13} shows how the SER varies with $k$ for $M = 50$ receptors, for the six choices of measurements.  Note that the SER curves for three choices of measurements --- measuring only one of the $C_{[2]}$, $C_{[3]}$ and $C_{[4]}$ --- are similar and we use one curve to represent their SER. The SER is lowest when we measured all the five species and this is expected. The figures show that: (1) If we measure $C_{[1]}$ only, the SER is comparable to measuring all five species; (2) If we only measure one of the $C_{[2]}$, $C_{[3]}$, $C_{[4]}$ and $C_{[5]}$, then the SER is much higher than measuring $C_{[1]}$ only. This suggests if we want to limit our observation to measuring only one species, then it is best to measure $C_{[1]}$ only. This conclusion is also consistent with the ligand-receptor binding with three binding sites. We have varied the value of $k$ and $M$, and similar results are obtained. Results are not shown due to space limit. 

We see that an application of the results of this paper is that it allows us to quickly derive the demodulation filters for different choices of molecular circuits and measurements. This allows us to evaluate the performance of these different design choices. 

%

\section{Conclusions}
\label{22}

This paper studies the demodulation of RSK signals. The derivation of demodulation filter requires a Bayesian filtering problem to be solved. This paper proposes a graphical method to derive the solution of this Bayesian filtering problem for any choice of molecular circuit at the receiver and any choice of measurements. We illustrate our proposed method with an example molecular circuit. We also present a numerical example to study the impact of the choice of measurements on communication performance.




\appendices
\section{Proof}
\label{app:proof}
This appendix aims to prove the validity of the general solution presented in Section \ref{sec:graph:alg}. We begin by describing the end-to-end CTMP model. 

\subsection{CTMP Model and Transition Probabilities}

\textcolor{black}{
The complete system which includes the transmitter, transmission medium and receiver can be modelled by a CTMP. In Section \ref{sec:e2e}, we presented a CTMP where the receiver uses a receptor with two binding sites. We will generalize the model to an arbitrary molecular circuit at the receiver.}

\textcolor{black}{
The state of the CTMP for the end-to-end system consists of the counts of all the chemical species in each voxel. We divide the elements of the state into two types: the observed and the unobserved species. The observed species are those that are found in the receiver and whose counts are made available for demodulation. 
Let the vectors $N_u(t)$ and $N_{o}(t)$ be respectively the counts of all the unobserved and observed species in the system at time $t$. For the two binding site example with only $C_{[1]}$ observed, $N_{o}(t)$ is $b_1(t)$ and $N_{u}(t)$ contains all the other molecular counts. Similarly, if both $C_{[1]}$ and $C_{[2]}$ are observed, then $N_{o}(t)$ is $[b_1(t) , b_2(t)]$, and $N_{u}(t)$ contains all the other molecular counts; note that for this case, $N_{u}(t)$ is the same as $N(t)$ in Section \ref{sec:e2e}. The state of CTMP is $[N_u(t), N_{o}(t)]$.}

\textcolor{black}{
We now specify the state transition probabilities from $[N_u(t), N_{o}(t)]$ to $[N_u(t+ \Delta t),  N_{o}(t+\Delta t)]$. We divide the state transitions into two types: (1) Transitions where $N_{o}(t+\Delta t) \neq N_{o}(t)$, and $N_{u}(t+\Delta t)$ can be the same or different from $N_{u}(t)$; (2) Transitions where $N_{o}(t+\Delta t) = N_{o}(t)$ and $N_{u}(t+\Delta t) \neq N_{u}(t)$. The first type of transitions is due to the occurrence of a chemical reaction that involves at least one observed species in the receiver. The second type of transitions include all other transitions, which are chemical reactions not involving any observed species and diffusion.}

\textcolor{black}{
Following the notation in Section \ref{sec:graph:alg}, the first type of transitions can be due to one of the $m_R$ reactions $R_1$, ..., $R_{m_R}$ that involve at least one observed species. Let us assume that an occurrence of $R_i$ causes the state change: $N_{u}(t+\Delta t) = N_{u}(t) + u_i$ and $N_{o}(t+\Delta t) = N_{o}(t) + o_i$. According to Equation \eqref{qw}, the rate of this reaction is: 
\begin{align}
\rho_i(N_{u}(t),N_{o}(t)) = \kappa_i   \Pi_{g = 1}^{m_O}  n_{O_g}(t)^{a_{i,g}} \Pi_{g = 1}^{m_U}  n_{U_g}(t)^{b_{i,g}} 
\end{align}
The state transition probability due to $R_i$ is:
\begin{align}
\mathbf{P}[ N_u(t + \Delta t) = N(t) + u_i ,  N_{o}(t+\Delta t) = N_{o}(t) +o_i | N_{u}(t),N_{o}(t)]  = \rho_i(N_{u}(t),N_{o}(t)) \Delta t \nonumber  \\ &
\end{align}
} 

\textcolor{black}{
For the second type of state transitions, if $N_{u}(t) = n_{(j)}$ and $N_{u}(t+\Delta t) = n_{(h)}$, the state transition probability is:
\begin{align}
{\bf P}[N_u(t+\Delta t) = n_{(h)} ,N_{o}(t+\Delta t)=N_{o}(t) |  N(t) = n_{(j)} ,N_{o}(t)]  = d_{hj}  \Delta t 
\label{eqn:tp:n}
\end{align}
where $d_{hj}$ denote the rate of transitions. We will not be specifying the precise form of  $d_{hj}$ because these parameters will be cancelled out during the derivation and will not appear in the final expression.}

\textcolor{black}{Finally the probability of no state transition, i.e. no change in the number of both the observed or unobserved species, is: 
\begin{align}
& {\bf P}[N_u(t+\Delta t) = N_u(t) , N_{o}(t+\Delta t) = N_{o}(t)|  N_u(t) = n_{(j)} ,N_{o}(t)]  = \nonumber  \\ &  1 - [\sum_{h \neq j} d_{hj}  +  \sum_{i=1}^{m_{R}} 
\rho_i(N_{u}(t),N_{o}(t))]  \; \Delta t 
\label{eqn:tp:n2}
\end{align}
}

\subsection{Bayesian Filtering Problem}

\textcolor{black}{Let ${\cal B}(t)$ denote the history of the observed species. Our aim is to determine $\mathbf{P} [N_{o}(t+\Delta t) | s, {\cal B}(t))$. By conditioning on the system state, we have:
\begin{align}
& \mathbf{P}[N_{o}(t + \Delta t) | s, {\cal B}(t) ]   \nonumber  \\ & 
	= \sum_{h}^{} \mathbf{P} [ N_u(t + \Delta t) = n_{(h)}, N_{o}(t + \Delta t) | s,  {\cal B}(t) ]    
	\nonumber  \\ & 
		= \sum_{h}^{} \sum_{j}^{} \mathbf{P} [ N_u(t + \Delta t) = n_{(h)} ,N_{o}(t + \Delta t) | s, N_u(t) = n_{(j)} ,  {\cal B}(t)  ]   \times \mathbf{P}  [N_u(t) = n_{(j)} | s , {\cal B}(t) ] 
		\nonumber  \\ & 
				= \sum_{h}^{} \sum_{j}^{} \mathbf{P} [ N_u(t + \Delta t) = n_{(h)} ,N_{o}(t + \Delta t) | s, N_u(t) = n_{(j)} , N_{o}(t) ]   \times \mathbf{P}  [N_u(t) = n_{(j)} | s , {\cal B}(t) ] 
	   \label{eqn:predictab}
\end{align}
}

\textcolor{black}{where we have used the Markov property $\mathbf{P} [ N_u(t + \Delta t) = n_{(h)} ,N_{o}(t + \Delta t) | s, N_u(t) = n_{(j)} , {\cal B}(t) ]$  = $\mathbf{P} [ N_u(t + \Delta t) = n_{(h)} ,N_{o}(t + \Delta t) | s, N_u(t) = n_{(j)} , N_{o}(t) ]$  to arrive at Equation \eqref{eqn:predictab}. We now focus on the term $\mathbf{P} [ N_u(t + \Delta t) = n_{(h)} ,N_{o}(t + \Delta t) | s, N_u(t) = n_{(j)} , N_{o}(t) ] $  in Equation \eqref{eqn:predictab}. This term is the state transition
probability and can be written as:
\begin{align}
&  \mathbf{P} [ N_u(t + \Delta t) = n_{(h)} ,N_{o}(t + \Delta t) | s, N_u(t) = n_{(j)} , N_{o}(t)  ] = \nonumber  \\ 
& \sum_{i=1}^{m_{R}}  \delta( N_{o}(t + \Delta t) = N_{o}(t) + o_h ) P_{i} 
+  \delta (N_{o}(t + \Delta t) = N_{o}(t)) P_{n} 
  \label{eqn:predictcb}
\end{align}
where 
\begin{align}
& P_{i} = \delta( n_{(h)} = n_{(j)} + u_i ) \rho_h(N_{u}(t),N_{o}(t)) \Delta t
  \label{eqn:predictcbsc} \\
& P_{n} = 
\begin{cases}
      d_{hj} \Delta t \; & \text{if}\ h \neq j \\
      1 - \sum_{h \neq j} d_{hj} \Delta t -  \sum_{i=1}^{m_{R}} \rho_i(N_{u}(t),N_{o}(t))  \Delta t\; & \text {if}\ h=j
    \end{cases}
  \label{eq:n}
\end{align}
}
\textcolor{black}{By substituting Eq. \eqref{eqn:predictcb} in Eq. \eqref{eqn:predictab}, we have:
\begin{align}
&\mathbf{P}[N_{o}(t + \Delta t) | s, {\cal B}(t) ] 
= \sum_{i=1}^{m_{R}} \delta( N_{o}(t + \Delta t) = N_{o}(t) + o_i ) Q_{i}
+  \delta (N_{o}(t + \Delta t) = N_{o}(t)) Q_{n} 
\label{eqn:predictcab}
\end{align}
where $Q_{i} = \sum_{h} \sum_{j} P_i  \mathbf{P}  [N_u(t) = n_{(j)} | s, {\cal B}(t) ]$ and $Q_{n} = \sum_{h} \sum_{j} P_n \mathbf{P}  [N_u(t) = n_{(j)} | s, {\cal B}(t) ]$. We will now derive the expressions for  $Q_{i}$ (for $i = 1,.., m_{R})$ and $Q_n$. For $Q_{i}$, we have: 
\begin{align}
& Q_{i} =  \sum_{h}^{} \sum_{j}^{} P_{i} \times \mathbf{P}  [N_u(t) = n_{(j)} | s,  {\cal B}(t)  ]  \nonumber  \\ 
&= \sum_{h}^{} \sum_{j}^{}    \delta(n_{(h)} = n_{(j)} + u_i ) \rho_i(N_{u}(t),N_{o}(t)) \Delta t  \mathbf{P}  [N_u(t) = n_{(j)} | s,  {\cal B}(t) ]  \nonumber  \\ 
&=  \sum_{j}^{}  \rho_i(N_{u}(t),N_{o}(t)) \Delta t  \mathbf{P}  [N_u(t) = n_{(j)} | s,  {\cal B}(t) ]  \nonumber  \\ 
&= E[ \rho_i(N_{u}(t),N_{o}(t)) | s,  {\cal B}(t) ] \Delta t
\label{eqn:predictcbc}
\end{align}
which means that $Q_{i}$ is the product of $\Delta t$ and the mean reaction rate of Reaction $i$ at time $t$ given the transmission symbol and past observations. Comparing the expression of $Q_{i}$ against that of $Q_{i.m}$ in Eq. \eqref{29}, we can identify $Q_i$ with the $Q_{i,m}$ in Section \ref{sec:graph:alg}. }

\textcolor{black} {For $Q_n$, we have:
\begin{align}
& Q_{n} =  \sum_{h}^{} \sum_{j}^{} P_{n} \times \mathbf{P}  [N_u(t) = n_{(j)} | s,  {\cal B}(t)   ] \nonumber \\
& = \sum_{h}^{} \sum_{j \neq h  }  d_{hj} \Delta t  \mathbf{P}  [N_u(t) = n_{(j)} | s,   {\cal B}(t)  ] \nonumber  \\
& + \sum_{j}^{} (1 - \sum_{h \neq j} d_{hj} \Delta t  -  \sum_{i=1}^{m_{R}}  \rho_i(N_{u}(t),N_{o}(t))   \Delta t) \mathbf{P}  [N_u(t) = n_{(j)} | s, {\cal B}(t)   ] \nonumber  \\
& = \sum_{j}^{} (1  -  \sum_{i=1}^{m_{R}} \rho_i(N_{u}(t),N_{o}(t)) \Delta t) \mathbf{P}  [N_u(t) = n_{(j)} | s, {\cal B}(t)  ] 
\nonumber  \\
&  \underbrace{ +  \sum_{h}^{} \sum_{j \neq h  }  d_{hj} \Delta t   \mathbf{P}  [N_u(t) = n_{(j)} | s,  {\cal B}(t) ]  - \sum_{j}^{} \sum_{h \neq j} d_{hj} \Delta t \mathbf{P}  [N_u(t) = n_{(j)} | s,  {\cal B}(t) ]}_{=0}
\nonumber  \\
 & = 1  -  \sum_{i=1}^{m_{R}} Q_{i} 
  \label{eqn:predictcbs}
\end{align}
Note that $Q_n$ here is identical to the expression of $Q_n$ in Section \ref{sec:graph:alg}. 
}



\ifCLASSOPTIONcaptionsoff
  \newpage
\fi



%

{\small
\bibliographystyle{IEEEtran}
\bibliography{nano,book}
}

\end{document}